\documentclass[12pt]{article}
\usepackage{pdproc,epsfig} 

\def\ov{\overline}

\def\nn{\nonumber}
\def\bea{\begin{eqnarray}}
\def\eea{\end{eqnarray}}
\def\beq{\begin{equation}}
\def\eeq{\end{equation}}
\def\bq{\begin{quote}}
\def\eq{\end{quote}}
\def\gappeq{\mathrel{\rlap {\raise.5ex\hbox{$>$}} {\lower.5ex\hbox{$\sim$}}}}
\def\lappeq{\mathrel{\rlap{\raise.5ex\hbox{$<$}} {\lower.5ex\hbox{$\sim$}}}}

\def\GeV{{\rm GeV}}
\def\PR{{\it Phys.~Rev.~}}
\def\PRL{{\it Phys.~Rev.~Lett.~}}
\def\NP{{\it Nucl.~Phys.~}}

\def\PL{{\it Phys.~Lett.~}}

\begin{document}
\vspace*{-1cm}
\phantom{hep-ph/0405048} 
\hfill{DFPD-04/TH/11}
\\
\phantom{hep-ph/0405048} 
\hfill{CERN-PH-TH/2004-079}
\vskip 2.0cm
\renewcommand{\thefootnote}{\alph{footnote}}
  
\title{MODELS OF NEUTRINO MASSES AND MIXINGS}

\author{GUIDO ALTARELLI}

\address{ CERN, Department of Physics, Theory Division \\  
CH-1211 Gen\`eve 23, Switzerland\\
 {\rm E-mail: guido.altarelli@cern.ch}}

  \centerline{\footnotesize and}

\author{FERRUCCIO FERUGLIO}

\address{Dipartimento di Fisica `G.~Galilei', Universit\`a di Padova and\\  
INFN, Sezione di Padova, Via Marzolo~8, I-35131 Padua, Italy\\
{\rm E-mail: feruglio@pd.infn.it}}

\abstract{We review theoretical ideas, problems and implications of neutrino
masses and mixing angles. We give a general discussion of schemes with three light neutrinos. Several specific examples are
analyzed in some detail, particularly those that can be embedded into grand unified theories.}
   
\normalsize\baselineskip=15pt

\section{Introduction}

There is by now convincing evidence, from the experimental study of atmospheric and solar neutrinos \cite{atmexp,sunexp,SNO,newSNO},
for the existence of at least two distinct frequencies of neutrino oscillations. This in turn implies non-vanishing
neutrino masses and a mixing matrix, in analogy with the quark sector and the CKM matrix. So apriori the study of masses
and mixings in the lepton sector should be considered at least as important as that in the quark sector. But actually
there are a number of features that make neutrinos especially interesting. In fact the smallness of neutrino masses is
probably related to the fact that
$\nu's$ are completely neutral (i.e. they carry no charge which is exactly conserved) and are Majorana particles with
masses inversely proportional to the large scale where lepton number (L) conservation is violated. Majorana masses can
arise from the see-saw mechanism \cite{seesaw}, in which case there is some relation with the Dirac masses, or from
higher-dimensional non-renormalizable operators which come from a different sector of the lagrangian density than any
other fermion mass terms. The relation with L non-conservation and the fact that the observed neutrino oscillation
frequencies are well compatible with a large scale for L non-conservation, points to a tantalizing connection with Grand
Unified Theories (GUT's). So neutrino masses and mixings can represent a probe into the physics at GUT energy scales and
offer a different perspective on the problem of flavour and the origin of fermion masses. There are also direct
connections with important issues in astrophysics and cosmology as for example baryogenesis through leptogenesis
\cite{leptgen} and the possibly non-negligible contribution of neutrinos to hot dark matter in the Universe.

Recently there have been new important experimental results that have considerably improved our knowledge. The SNO
experiment has confirmed that the solar neutrino deficit is due to neutrino oscillations and not to a flaw in our modeling
of the sun \cite{SNO}: the total neutrino flux is in agreement with the solar model but only about one third arrives on Earth as
$\nu_e$ while the remaining part consists of other kinds of active neutrinos, presumably $\nu_{\mu}$ and $\nu_{\tau}$. The
allowed amount of sterile neutrinos is strongly constrained.  The KamLAND experiment has established that 
$\ov{\nu_e}$ from reactors show oscillations over an average distance of about 180 Km which are perfectly compatible with
the frequency and mixing angle corresponding to one of the solutions of the solar neutrino problem (the Large Angle
(LA) solution) \cite{kam}. Thus the results from solar neutrinos have been reproduced and improved by a terrestrial experiment. Also
the coincidence of the frequency for neutrinos from the sun and for antineutrinos from reactors is consistent with the
validity of CPT invariance. The validity of this symmetry had been questioned because of the puzzling LSND claim of a signal
that could indicate a third distinct oscillation frequency (hence implying either more than three light neutrinos or CPT
violation). In September '03 new results have been published by the SNO Collaboration \cite{newSNO}, obtained after adding salt to their heavy water detector in order to increase the sensitivity to the neutral current channels. The previous results have been confirmed with increased accuracy. The allowed region for the LA solution has been further restricted with the elimination of the upper region in $\Delta m^2_{12}$. Of great importance have also been the first results from WMAP \cite{wmap} on the cosmic radiation background. The related determination of
cosmological parameters, in combination with other measurements, leads to an upper limit on the cosmological neutrino
density $\Omega_{\nu} \lappeq 0.015$. This is a very important result that indicates that neutrinos are not a major
component of the dark matter in the Universe. For three degenerate neutrinos the WMAP limit implies an upper bound on the
common mass given by $m_{\nu}< 0.23$ eV \cite{wmap}. Given the priors that are assumed for this determination, i.e. a definite
cosmological model, a 2-digit value for the bound is not to be taken too seriously \cite{hann}. Still the quoted value is about an order
of magnitude smaller than the bound from tritium beta decay and of the same order of the upper bound on the Majorana mass that
fixes the rate of neutrinoless double beta decay.   

In spite of this progress there are many alternative models of neutrino masses \cite{revs}. This variety is mostly due to the
considerable experimental ambiguities that still exist. One first missing input is the absolute scale of neutrino masses:
neutrino oscillations only determine mass squared differences. For atmospheric neutrinos $\Delta
m^2_{atm}~\sim~2.6~10^{-3}$ eV$^2$ while for solar neutrinos $\Delta m^2_{sol}~\sim~7~10^{-5}$ eV$^2$. Another key missing 
quantity is the value of
the third mixing angle $s_{13}$ on which only a bound is known, $s_{13}< 0.22$. Then it is essential to know whether
the LSND signal \cite{LSND}, which has not been confirmed by KARMEN
\cite{karmen} and is currently being double-checked by MiniBoone
\cite{MBoone}, will be confirmed or will be excluded.  If LSND is right we probably
need at least four light neutrinos; if not we can do with only the three known ones. 

Here we will briefly summarize the main categories of neutrino mass models, discuss their respective advantages and
difficulties and give a number of examples. We illustrate how forthcoming experiments can discriminate among the various
alternatives. We will devote a special attention to  a comprehensive discussion in a GUT framework of neutrino masses
together with all other fermion masses. This is for example possible in models based on SU(5)$\times$ U(1)$_{\rm F}$ or
on SO(10) (we always consider SUSY GUT's) \cite{us4,barr}.

\section{Basic Formulae and Data for Three-Neutrino Mixing}

We assume in the following that the LSND signal \cite{LSND}, will not be confirmed  so that there are only two distinct
neutrino oscillation frequencies, the atmospheric and the solar frequencies. These two can be reproduced with the known
three light neutrino species (for more than three neutrinos see, for example, ref. \cite{4nu}). 

Neutrino oscillations are due to a misalignment between the flavour basis, $\nu'\equiv(\nu_e,\nu_{\mu},\nu_{\tau})$, where
$\nu_e$ is the partner of the mass and flavour eigenstate $e^-$ in a left-handed (LH) weak isospin SU(2) doublet (similarly
for 
$\nu_{\mu}$ and $\nu_{\tau})$) and the mass eigenstates $\nu\equiv(\nu_1, \nu_2,\nu_3)$ \cite{pon,lee}: 
\beq
\nu' =U \nu~~~,
\label{U}
\eeq  where $U$ is the unitary 3 by 3 mixing matrix. Given the definition of $U$ and the transformation properties of the
effective light neutrino mass matrix $m_{\nu}$:
\bea 
\label{tr} {\nu'}^T m_{\nu} \nu'&= &\nu^T U^T m_\nu U \nu\\ \nonumber  U^T m_{\nu} U& = &{\rm
Diag}\left(m_1,m_2,m_3\right)\equiv m_{diag}~~~,
\eea  we obtain the general form of $m_{\nu}$ (i.e. of the light $\nu$ mass matrix in the basis where the charged lepton
mass is a diagonal matrix):
\beq  m_{\nu}=U^* m_{diag} U^\dagger~~~.
\label{gen}
\eeq  The matrix $U$ can be parameterized in terms of three mixing angles $\theta_{12}$,
$\theta_{23}$ and $\theta_{13}$ ($0\le\theta_{ij}\le \pi/2$)  and one phase $\varphi$ ($0\le\varphi\le 2\pi$) \cite{cab},
exactly as for the quark mixing matrix $V_{CKM}$. The following definition of mixing angles can be adopted:
\beq  U~=~ 
\left(\matrix{1&0&0 \cr 0&c_{23}&s_{23}\cr0&-s_{23}&c_{23}     } 
\right)
\left(\matrix{c_{13}&0&s_{13}e^{i\varphi} \cr 0&1&0\cr -s_{13}e^{-i\varphi}&0&c_{13}     } 
\right)
\left(\matrix{c_{12}&s_{12}&0 \cr -s_{12}&c_{12}&0\cr 0&0&1     } 
\right)
\label{ufi}
\eeq  where $s_{ij}\equiv \sin\theta_{ij}$, $c_{ij}\equiv \cos\theta_{ij}$.  In addition, if $\nu$ are Majorana particles,
we have the relative phases among the Majorana masses
$m_1$, $m_2$ and $m_3$. If we choose $m_3$ real and positive, these phases are carried by $m_{1,2}\equiv\vert m_{1,2}
\vert e^{i\phi_{1,2}}$
\cite{frsm}.  Thus, in general, 9 parameters are added to the SM when non-vanishing neutrino masses are included: 3
eigenvalues, 3 mixing angles and 3 CP  violating phases.

In our notation the two frequencies, $\Delta m^2_{I}/4E$ $(I=sun,atm)$, are parametrized in terms of the $\nu$ mass
eigenvalues by 
\beq
\Delta m^2_{sun}\equiv \vert\Delta m^2_{12}\vert ,~~~~~~~
\Delta m^2_{atm}\equiv \vert\Delta m^2_{23}\vert~~~.
\label{fre}
\eeq   where $\Delta m^2_{12}=\vert m_2\vert^2-\vert m_1\vert^2 > 0$ and $\Delta m^2_{23}= m_3^2-\vert m_2\vert ^2$. The
numbering 1,2,3 corresponds to our definition of the frequencies and in principle may not coincide with the ordering from
the lightest to the heaviest state. From experiment, see table \ref{tab01} \cite{globfit}, we know that $s_{13}$ is small,  according to
CHOOZ, $s_{13}<0.22$ (3$\sigma$) \cite{chooz}.
%%%%%%%%%%%%%%%%%%%%%%%%%%%%%%%%%%%%%%%%%%%%%%%%%%%%%%%%%%%%%%%%%%%%%%%%%%%%%
\vspace{0.1cm}
\begin{table}[!t]
\caption{Square mass differences and mixing angles.}
\label{tab01}
\vspace{0.4cm}
\begin{center}
\begin{tabular}{|c|c|c|c|}   
\hline  
& & & \\   
&{\tt lower limit} & {\tt best value} & {\tt upper limit}\\
&($3\sigma$)& & ($3\sigma$)\\
\hline
& & & \\
$(\Delta m^2_{sun})_{\rm LA}~(10^{-5}~{\rm eV}^2)$ & 5.4& 6.9&9.5\\
& & & \\
\hline
& & & \\
$\Delta m^2_{atm}~(10^{-3}~{\rm eV}^2)$ & 1.4& 2.6& 3.7\\
& & & \\
\hline
& & & \\
$\sin^2\theta_{12}$ & 0.23 & 0.30 &0.39\\
& & & \\
\hline
& & & \\
$\sin^2\theta_{23}$ & 0.31 & 0.52 &0.72\\
& & & \\
\hline
& & & \\
$\sin^2\theta_{13}$ & 0 & 0.006 &0.054\\
& & & \\
\hline
\end{tabular} 
\end{center}
\end{table}
% mettere nel testo \cite{sunfit,atmfit}
Atmospheric neutrino oscillations mainly depend on $(\Delta m_{atm}^2,
\theta_{23},\theta_{13})$, while solar oscillations are controlled by
$(\Delta m_{sol}^2,\theta_{12},\theta_{13})$. Therefore, in the ideal limit  of exactly vanishing $s_{13}$, the solar and
atmospheric oscillations decouple and depend on two separate sets of two-flavour parameters.    For atmospheric neutrinos we
have $c_{23}\sim s_{23}\sim 1/\sqrt{2}$,  corresponding to nearly maximal mixing. Oscillations of muon neutrinos
into  tau neutrinos are favoured over oscillations into sterile neutrinos ($\nu_s$). The conversion probability and the
zenith angular distribution of high-energy  muon neutrinos are sensitive to matter effects, which distinguish 
$\nu_\tau$ from $\nu_s$. Moreover, for conversion of $\nu_\mu$ into pure $\nu_s$, neutral current events would become
up/down asymmetric. In both cases data strongly disfavour the pure sterile case. Oscillations into $\nu_\tau$ are also
indirectly supported by a SK data sample that can be interpreted in terms of enriched $\tau$-like charged-current events. 
The sterile component of the neutrino participating in atmospheric oscillations should amount to less than 0.25 at 90\%
C.L. Disappearance of laboratory-produced muon neutrinos has also been confirmed within expectations by the K2K
experiment. 
 
The only surviving solution to the solar neutrino problem after KamLAND and SNO-salt results is LA \cite{sunfit,atmfit,globfit}, with $\Delta
m_{sol}^2\approx 7\cdot 10^{-5}$ eV$^2$ and $\sin^2\theta_{12}\approx 0.3$.  Before KamLAND the interpretation of solar neutrino data in terms of oscillations required the
knowledge of the Boron neutrino flux, $f_B$. For instance, charged and neutral current data from SNO are sensitive,
respectively, to $f_B \langle P_{ee}\rangle$ and to $f_B \langle \sum_a P_{ea} \rangle$ $(a=e,\nu,\tau)$, where
$\langle P_{ef} \rangle$ denotes the appropriately averaged conversion probability from $\nu_e$ to $\nu_f$. KamLAND
\cite{kam} provides a direct measurement of $\langle P_{ee}\rangle$. Beyond the impact on the oscillation parameters and a check that
the solar standard model works well ($f_B=(1.00\pm0.06)\times5.05\cdot 10^{6}~cm^{-2}  sec^{-1}$), the comparison among
these experiments shows that the  conversion of Boron solar neutrinos into sterile neutrinos is compatible with zero.
Therefore, the LSND indication for a third  oscillation frequency associated to one or more  sterile neutrinos is not
supported by any other experiment, at the moment. Now, after KamLAND, also the possibility that such a frequency originates
from a CPT violating neutrino spectrum \cite{cpt} has no independent support. Data from solar neutrino experiments and 
KamLAND, involving,
respectively, electron neutrinos and electron antineutrinos, are compatible with a CPT invariant spectrum. 

If we take maximal $s_{23}$ and keep only linear terms in $u=  s_{13}e^{i\varphi}$, from experiment we find the following
structure of the
$U_{fi}$ ($f=e$,$\mu$,$\tau$, $i=1,2,3$) mixing matrix, apart from sign convention redefinitions: 
\beq  U_{fi}= 
\left(\matrix{ c_{12}&s_{12}&u \cr  -(s_{12}+c_{12}u^*)/\sqrt{2}&(c_{12}-s_{12}u^*)/\sqrt{2}&1/\sqrt{2}\cr
(s_{12}-c_{12}u^*)/\sqrt{2}&-(c_{12}+s_{12}u^*)/\sqrt{2}&1/\sqrt{2}     } 
\right) ~~~~~,
\label{ufi1}
\eeq  where $\theta_{12}$ is close to $\pi/6$ (for  $s_{12}=1/\sqrt{3}$ and $u=0$ we have the so-called tri-bimaximal mixing pattern \cite{bitri}, with the entries in the second column all equal to $1/\sqrt{3}$ in absolute value). Given the observed frequencies and  our notation in eq. (\ref{fre}), there
are three possible patterns of mass eigenvalues:
\bea {\tt{Degenerate}}& : & |m_1|\sim |m_2| \sim |m_3|\gg |m_i-m_j|\nonumber\\ {\tt{Inverted~hierarchy}}& : & |m_1|\sim
|m_2| \gg |m_3| \nonumber\\ {\tt{Normal~hierarchy} }& : & |m_3| \gg |m_{2,1}|
\label{abc}
\eea  
Models based on all these patterns have been proposed and studied and all are in fact viable at present. In the
following we will first discuss neutrino masses in general and, in particular, Majorana neutrinos. Then we recall the
existing constraints on the absolute scale of neutrino masses. We then discuss the importance of
neutrinoless double beta decay that, if observed, would confirm the Majorana nature of neutrinos. Also the knowledge of
the rate of this process could discriminate among the possible patterns of neutrino masses in (\ref{abc}). The
possible importance of heavy Majorana neutrinos for the explanation of baryogenesis through leptogenesis in the early
Universe will be briefly discussed. We finally review the phenomenology of neutrino mass models based on the three
spectral patterns in (\ref{abc}) and the respective advantages and problems.

\section{Neutrino Masses and Lepton Number Violation}

Neutrino oscillations imply neutrino masses which in turn demand either the existence of right-handed (RH) neutrinos
(Dirac masses) or lepton number L violation (Majorana masses) or both. Given that neutrino masses are certainly extremely
small, it is really difficult from the theory point of view to avoid the conclusion that L conservation must be violated.
In fact, in terms of lepton number violation the smallness of neutrino masses can be explained as inversely proportional
to the very large scale where L is violated, of order $M_{GUT}$ or even $M_{Pl}$.

Once we accept L non-conservation we gain an elegant explanation for the smallness of neutrino masses. If L is not
conserved, even in the absence of heavy RH neutrinos, Majorana masses for neutrinos can be generated by dimension five
operators \cite{weinberg} of the form 
\beq 
O_5=\frac{(H l)^T_i \lambda_{ij} (H l)_j}{\Lambda}+~h.c.~~~,
\label{O5}
\eeq  
with $H$ being the ordinary Higgs doublet, $l_i$ the SU(2) lepton doublets, $\lambda$ a matrix in  flavour space,
$\Lambda$ a large scale of mass, of order $M_{GUT}$ or $M_{Pl}$ and a charge conjugation matrix $C$
between the lepton fields is understood. 
Neutrino masses generated by $O_5$ are of the order
$m_{\nu}\approx v^2/\Lambda$ for $\lambda_{ij}\approx {\rm O}(1)$, where $v\sim {\rm O}(100~\GeV)$ is the vacuum
expectation value of the ordinary Higgs. 

We consider that the existence of RH neutrinos $\nu^c$ is quite plausible because all GUT groups larger than SU(5) require
them. In particular the fact that $\nu^c$ completes the representation 16 of SO(10): 16=$\bar 5$+10+1, so that all
fermions of each family are contained in a single representation of the unifying group, is too impressive not to be
significant. At least as a classification group SO(10) must be of some relevance. Thus in the following we assume that
there are both
$\nu^c$ and L non-conservation. With these assumptions the see-saw mechanism \cite{seesaw} is possible.  Also to fix
notations we recall that in its simplest form it arises as follows. Consider the SU(3) $\times$ SU(2) $\times$ U(1)
invariant Lagrangian giving rise to Dirac and $\nu^c$ Majorana masses (for the time being we consider the $\nu$
(versus $\nu^c$) Majorana  mass terms as comparatively negligible):
\beq 
{\cal L}=-{\nu^c}^T y_\nu (H l)+\frac{1}{2}
{\nu^c}^T M \nu^c +~h.c.
\label{lag}
\eeq  
The Dirac mass matrix $m_D\equiv y_\nu v/\sqrt{2}$, originating from electroweak symmetry breaking,  is, in general,
non-hermitian and non-symmetric, while the Majorana mass matrix $M$ is symmetric,
$M=M^T$. We expect the eigenvalues of $M$ to be of order $M_{GUT}$ or more because $\nu^c$ Majorana masses are
SU(3)$\times$ SU(2)$\times$ U(1) invariant, hence unprotected and naturally of the order of the cutoff of the low-energy
theory.  Since all $\nu^c$ are very heavy we can integrate them away.  For this purpose we write down the equations of
motion for $\nu^c$ in the static limit, $i.e.$ neglecting their kinetic terms:
\beq  -\frac{\partial {\cal L}}{\partial\nu^c}=y_\nu (H l)- M \nu^c= 0~~~.
\label{eulag}
\eeq  {}From this, by solving for $\nu^c$, we obtain:
\beq
\nu^c= M^{-1} y_\nu (H l)~~~.
\label{R}
\eeq  We now replace in the lagrangian, eq. (\ref{lag}), this expression for $\nu^c$ and we get the operator $O_5$ of eq.
(\ref{O5}) with
\beq 
\frac{2 \lambda}{\Lambda}=-y_\nu^T M^{-1} y_\nu ~~~~~,
\eeq and the resulting neutrino mass matrix reads:
\beq  m_{\nu}=m_D^T M^{-1}m_D~~~.
\eeq  This is the well known see-saw mechanism result \cite{seesaw}: the light neutrino masses are quadratic in the Dirac
masses and inversely proportional to the large Majorana mass.  If some $\nu^c$ are massless or light they would not be
integrated away but simply added to the light neutrinos. Notice that the above results hold true for any number
$n$ of heavy neutral fermions 
$R$ coupled to the 3 known neutrinos. In this more general case $M$ is an $n$ by $n$ symmetric matrix and the coupling
between heavy and light fields is described by the rectangular $n$ by 3 matrix $m_D$.  Note that for
$m_{\nu}\approx \sqrt{\Delta m^2_{atm}}\approx 0.05$ eV and 
$m_{\nu}\approx m_D^2/M$ with $m_D\approx v
\approx 200~GeV$ we find $M\approx 10^{15}~GeV$ which indeed is an impressive indication for
$M_{GUT}$.

If additional non-renormalizable contributions to $O_5$, eq. (\ref{O5}), are comparatively non-negligible, they should
simply be added.  For instance in SO(10) or in left-right extensions of the SM, an SU(2)$_L$ triplet can couple to 
lepton doublets and may induce a sizeable contribution to neutrino masses. At the level of the 
low-energy effective theory, such contribution is still described by the operator $O_5$ of eq. (\ref{O5}),
obtained by integrating out the heavy SU(2)$_L$ triplet (which also acquires a VEV due to its
coupling to the Higgs doublets). This contribution is called type II
to be distinguished from that obtained by the exchange of RH neutrinos (type I).
After elimination of the heavy fields, at the level of the effective low-energy theory, the
two types of terms are equivalent. In particular they have identical transformation properties under a chiral change of
basis in flavour space. The difference is, however, that in the see-saw mechanism, the Dirac matrix
$m_D$ is presumably related to ordinary fermion masses because they are both generated by the Higgs mechanism and both
must obey GUT-induced constraints. Thus if we assume the see-saw mechanism in its simplest  type I version,
more constraints are implied.

\section{Importance of Neutrinoless Double Beta Decay}

Oscillation experiments do not provide information about the absolute neutrino spectrum and cannot distinguish between
pure Dirac and Majorana neutrinos. From the endpoint of tritium beta decay spectrum we have  an absolute upper limit of
2.2 eV (at 95\% C.L.) on the mass of electron  antineutrino \cite{tritium}, which, combined with the observed oscillation
frequencies under the assumption of three CPT-invariant light neutrinos, represents also an upper bound on the masses of
the other active neutrinos. Complementary information on the sum of neutrino masses is also provided by the galaxy power
spectrum combined with measurements of the cosmic  microwave background anisotropies. According to the recent analysis of
the WMAP collaboration \cite{wmap}, 
$\sum_i \vert m_i\vert < 0.69$ eV (at 95\% C.L.).  More conservative analyses \cite{hann} give $\sum_i \vert
m_i\vert < 1.01$ eV, still much more restrictive than the laboratory bound.

The discovery of $0\nu \beta \beta$ decay would be very important because it would establish lepton number violation and
the Majorana nature of $\nu$'s, and provide direct information on the absolute
scale of neutrino masses.
As already mentioned the present limit from $0\nu \beta \beta$ is $\vert m_{ee}\vert< 0.2$ eV or to be
more conservative
$\vert m_{ee}\vert < (0.3\div 0.5)$ eV \cite{0nubblim,bilenky}. 
Note, however, that the WMAP limit implies for 3 degenerate
$\nu$'s $|m|< 0.23$ eV and, taken at face value, this limit would pose a direct constraint on $m_{ee}$. 

It is interesting to see
what is the level at which a signal can be expected or at least not excluded in the different classes of models in
(\ref{abc})
\cite{0nubb,fsv}.  The quantity which is bound by experiments
is the 11 entry of the
$\nu$ mass matrix, which in general, from eqs. (\ref{tr}) and (\ref{ufi}), is given by :
\beq 
\vert m_{ee}\vert~=\vert(1-s^2_{13})~(m_1 c^2_{12}~+~m_2 s^2_{12})+m_3 e^{2 i\phi} s^2_{13}\vert~~~,
\label{3nu1gen}
\eeq
For 3-neutrino models
with degenerate, inverse hierarchy or normal hierarchy mass patterns, starting from this general formula it is simple to
derive the following bounds.
\begin{itemize}
\item[a)]  Degenerate case. If $|m|$ is the common mass and we take $s_{13}=0$, which is a safe
approximation in this case, because $|m_3|$ cannot compensate for the smallness of $s_{13}$, we have
$m_{ee}\sim |m|(c_{12}^2\pm s_{12}^2)$.  Here the phase ambiguity has been reduced to a sign ambiguity which is sufficient
for deriving bounds.  So, depending on the sign we have
$m_{ee}=|m|$ or
$m_{ee}=|m|cos2\theta_{12}$. We conclude that in  this case $m_{ee}$ could be as large as the present experimental limit
but should be at least of
order $O(\sqrt{\Delta m^2_{atm}})~\sim~O(10^{-2}~ {\rm eV})$ unless the solar angle is practically maximal, in which case
the minus sign option can be arbitrarily small. But the experimental 2-$\sigma$ range of the solar angle does not
favour a cancellation by more than a factor of 3.
\item[b)]  Inverse hierarchy case. In this case the same approximate formula $m_{ee}=|m|(c_{12}^2\pm s_{12}^2)$ holds 
because $m_3$ is small and $s_{13}$ can be neglected. The difference is that here we know that $|m|\approx 
\sqrt{\Delta m^2_{atm}}$ so that $\vert m_{ee}\vert<\sqrt{\Delta m^2_{atm}}~\sim~0.05$ eV. At the same time,
since a full cancellation between the two contributions cannot take place, we expect 
$\vert m_{ee}\vert > 0.01$ eV.

\item[c)]  Normal hierarchy case. Here we cannot in general neglect the $m_3$ term. However in this case $\vert
m_{ee}\vert~\sim~
\vert\sqrt{\Delta m^2_{sun}}~ s_{12}^2~\pm~\sqrt{\Delta m^2_{atm}}~ s_{13}^2\vert$ and we have the bound 
$\vert m_{ee}\vert <$ a few $10^{-3}$ eV.
\end{itemize}

Recently evidence for $0\nu \beta \beta$ was claimed in ref. \cite{kla} at the 4.2-$\sigma$ level corresponding to
$\vert m_{ee}\vert\sim (0.2\div 0.6)~ {\rm eV}$ ($(0.1\div 0.9)~ {\rm eV}$ in a more conservative estimate of the involved nuclear matrix elements). If confirmed this would rule out cases b) and c) and point to case a) or to models
with more than 3 neutrinos. We recall that further contributions to $0\nu \beta \beta$ transition amplitudes
might occur in models with additional L violating interactions, such as R-parity breaking supersymmetry. 

\section{Baryogenesis via Leptogenesis from Heavy $\nu^c$ Decay}

In the Universe we observe an apparent excess of baryons over antibaryons. It is appealing that one can explain the
observed baryon asymmetry by dynamical evolution (baryogenesis) starting from an initial state of the Universe with zero
baryon number.  For baryogenesis one needs the three famous Sakharov conditions: B violation, CP violation and no thermal
equilibrium. In the history of the Universe these necessary requirements can have occurred at different epochs. Note
however that the asymmetry generated by one epoch could be erased at following epochs if not protected by some dynamical
reason. In principle these conditions could be verified in the SM at the electroweak phase transition. B is violated by
instantons when kT is of the order of the weak scale (but B-L is conserved), CP is violated by the CKM phase and
sufficiently marked out-of- equilibrium conditions could be realized during the electroweak phase transition. So the
conditions for baryogenesis  at the weak scale in the SM superficially appear to be present. However, a more quantitative
analysis
\cite{rev} shows that baryogenesis is not possible in the SM because there is not enough CP violation and the phase
transition is not sufficiently strong first order, unless
$m_H<80~{\rm GeV}$, which is by now completely excluded by LEP. In SUSY extensions of the SM, in particular in the MSSM,
there are additional sources of CP violation and the bound on $m_H$ is modified by a sufficient amount by the presence of
scalars with large couplings to the Higgs sector, typically the s-top. What is required is that
$m_h\sim 80-110~{\rm GeV}$, a s-top not heavier than the top quark and, preferentially, a small
$\tan{\beta}$. But also this possibility has by now become at best marginal with the results from LEP2.

If baryogenesis at the weak scale is excluded by the data it can occur at or just below the GUT scale, after inflation.
But only that part with
$|{\rm B}-{\rm L}        |>0$ would survive and not be erased at the weak scale by instanton effects. Thus baryogenesis at
$kT\sim 10^{10}-10^{15}~{\rm GeV}$ needs B-L violation at some stage like for $m_\nu$ if neutrinos are Majorana particles.
The two effects could be related if baryogenesis arises from leptogenesis then converted into baryogenesis by instantons
\cite{leptgen}. Recent results on neutrino masses are compatible with this elegant possibility \cite{kaol}. Thus the case
of baryogenesis through leptogenesis has been boosted by the recent results on neutrinos \cite{leptog}.

In leptogenesis the baryon asymmetry is produced by the out of equilibrium, CP and L-violating decays
of heavy right-handed neutrinos $\nu^c$. In the simplest cases the spectrum of RH neutrinos
is assumed to be hierarchical, $M_1\ll M_{2,3}$ and the mechanism is dominated by the lightest state, $\nu^c_1$. 
Thus the expected baryon asymmetry is proportional to the product $\epsilon_1 \delta$
between the CP decay asymmetry
\beq
\epsilon_1=\frac{\Gamma(\nu^c_1\to l)-\Gamma(\nu^c_1\to \bar{l})}{\Gamma(\nu^c_1\to l)+\Gamma(\nu^c_1\to \bar{l})}
~~~~~,
\eeq
and the efficiency factor $\delta\le 1$, the fraction of the produced asymmetry that survives after
$\nu^c_1$ decay. 
To keep $\delta$ close to 1 and avoid washing out the developed lepton asymmetry, L violating interactions
should be sufficiently weak to stay out of equilibrium when $\nu_1^c$ decays. 
This in turns happens when the lifetime of $\nu^c_1$ exceeds the age of the Universe,
which is expressed by the condition \cite{m1t}:
\beq
\tilde{m_1}\equiv\frac{(m_D m_D^\dagger)_{11}}{M_1}\le 10^{-3}~{\rm eV}~~~~~,
\label{ooe}
\eeq
with ${\rm min}\{|m1|,|m2|,|m3|\}\le\tilde{m_1}$. This condition does not provide yet
an absolute bound on neutrino masses since, in realistic simulations, the observed baryon asymmetry
is achieved for $\delta$ less than one. 
For hierarchical RH neutrino masses the asymmetry $\epsilon_1$ is given by:
\beq
\epsilon_1\approx -\frac{3\sigma}{8 \pi}\frac{1}{(y_\nu y_\nu^\dagger)_{11}} \sum_{j=2,3} 
Im\left[(y_\nu y_\nu^\dagger)_{1j}^2\right]
\frac{M_1}{M_j}~~~~~,
\label{eps1}
\eeq
($\sigma=1$ in the supersymmetric case and 1/2 in the non-supersymmetric one) 
which gives rise to the Davidson-Ibarra (DI) bound \cite{dibo}:
\beq
\vert\epsilon_1\vert\le \frac{3\sigma}{8 \pi}\frac{M_1}{\langle H^0\rangle^2}
({\rm max}\{|m1|,|m2|,|m3|\}-{\rm min}\{|m1|,|m2|,|m3|\})~~~~~,
\label{dib}
\eeq
where $\langle H^0\rangle$ denotes the VEV of the Higgs doublet giving mass to the up quarks.
Several interesting constraints on the neutrino spectrum emerge from the above discussion.
First of all, in the assumed limit $M_{2,3}\gg M_1$, the CP asymmetry vanishes for a completely
degenerate light neutrino spectrum (of course this limit is rather unnatural in the see-saw context). 
Moreover, depending on the details of the assumed cosmological scenario
which fixes the initial abundance of RH neutrinos, the DI bound translates into a lower bound
on the lightest RH neutrino mass $M_1$, ranging from approximately $10^7$ GeV to $10^9$ GeV \cite{dibo,mbounds}.
Finally, an absolute upper bound on light neutrino masses can be derived. Indeed, if we enhance the absolute
scale of light neutrino masses, on the one hand we depart from the out-of-equilibrium condition
of eq. (\ref{ooe}) and on the other hand the DI bound, proportional to $\Delta m^2_{atm}/m_3$, 
becomes more and more stringent. Quantitative studies \cite{mbounds} show that
\beq
|m_i|<(0.12\div 0.15)~{\rm eV}~~~~~.
\eeq
It should be stressed that these bounds hold under the assumption of strict hierarchy in the RH
neutrino sector. Indeed, by relaxing the assumption $M_{2,3}\gg M_1$, there are important corrections
to the expression of $\epsilon_1$ in eq. (\ref{eps1}), which may receive a resonant enhancement and which do not vanish any longer
for degenerate light neutrinos. For a moderate degeneracy among RH neutrinos, the DI bound is violated
and $M_1$ can be considerably lower than $10^7-10^9$ GeV \cite{ham}. At the same time, the enhanced CP violating asymmetry
allows a successful leptogenesis even for light neutrino masses around the eV scale \cite{ham}.
In any case it is impressive that the resulting range of neutrino masses is fully
consistent with the results on neutrino oscillations.

\section{Degenerate Neutrinos}

For degenerate neutrinos the average $m^2$ is much larger than the splittings. At first sight the degenerate case is the
most appealing: the observation of nearly maximal atmospheric neutrino mixing and the more recent result that also
the solar mixing is large suggests that all $\nu$ masses are nearly degenerate. We shall see that this possibility has
become less attractive with the recent new experimental information. 

It is clear that in the degenerate case
the most likely origin of $\nu$ masses is from some dimension 5 operators $(H l)^T_i\lambda_{ij}(H l)_j/\Lambda$ not
related to the see-saw mechanism 
$m_{\nu}=m^T_DM^{-1}m_D$. In fact we expect the $\nu$ Dirac mass $m_D$ not to be degenerate like  for all other fermions
and a conspiracy  to reinstate a nearly perfect degeneracy between $m_D$ and $M$, which arise from completely different
physics, looks very unplausible (see, however,
\cite{jeza}). Thus in degenerate models, in general, there is no direct relation with Dirac masses of quarks and leptons
and the possibility of a simultaneous description of all fermion masses within a grand unified theory is more remote
\cite{fri,wett}.

The degeneracy of neutrinos should be guaranteed by some slightly broken symmetry. Models based on discrete or continuous
symmetries have been proposed. For example in the models of ref. \cite{BHKR,AK} the symmetry is SO(3): in the unbroken limit
neutrinos are degenerate and charged leptons are massless. When the symmetry is broken the charged lepton masses are much
larger than neutrino splittings because the former are first order while the latter are second order in the electroweak
symmetry breaking. In this kind of models the mixing angles are completely undetermined in the symmetric phase
and they originate only in the spontaneously broken phase from a misalignment between the symmetry breaking terms
for neutrinos and charged leptons.

In principle, when considering models with degenerate masses we must keep in mind that radiative corrections can modify mass 
splittings and mixing angles in the running from the high scale where neutrino masses are determined at the fundamental 
level (i.e. the heavy Majorana mass $M$ or $M_{GUT}$) down to the electroweak scale \cite{radcor}. These running effects can be 
evaluated by renormalisation group techniques. The effects depend on the parameters 
$A_{ab}=(m_{a}+ m_{b})/( m_{a}-m_{b})$ and are, with good accuracy, proportional to 
$\epsilon = y_\tau^2/(16\pi^2) \log(M/m_Z)$, where $ y_\tau$ is the $\tau$-lepton Yukawa coupling. The value of 
$\epsilon$ is around $10^{-5}$ in the SM and larger by a factor $1+\tan^2{\beta}$ in the MSSM. The corrections are negligible 
for $|A_{ab} \epsilon| \lappeq 0(1)$. When some of these quantities are large a rapid transition takes place towards a fixed 
point configuration of mixing angles that does not correspond to the observed pattern. In practice, given the present upper 
bounds on the degenerate neutrino common mass $m_0$ and the LA value of $\Delta m^2_{sol}$, the corrections are always 
negligible in the SM and can only become sizable in the MSSM if $\tan^2{\beta}$ is large, $m_0$ is close to its absolute 
upper bound and $m_1$ and $m_2$ (those entering in the smallest difference $\Delta m^2_{sol}$) are nearly equal in absolute 
value and sign. Note that these remarks do not include the effects from thresholds that could affect running in a significant 
way, depending on the details of the heavy and/or light particle spectrum.

The upper limit on the common
value $|m|$ becomes particularly stringent if one adopts the cosmological WMAP bound $|m|< 0.23$ eV \cite{wmap}
(or the more conservative one $|m|< 0.34$ eV \cite{hann}). The more direct
laboratory limit from tritium beta decay is $|m|< 2.2$ eV \cite{tritium}. 
In past years degenerate models with $\nu$ masses as
large as
$|m| \sim (1\div 2)$ 
eV were considered with the perspective of a large fraction of hot dark matter in the universe. In this
case, however, the existing limit \cite{0nubblim} on the absence of 
$0\nu\beta\beta$ ($\vert m_{ee}\vert< 0.2$ eV or to be more conservative $\vert m_{ee}\vert< (0.3\div 0.5)$ eV) implies \cite{gg,bilenky}
approximate double maximal mixing (bimixing) for solar and atmospheric neutrinos. As discussed in sect. 4 , for $|m|\gg
m_{ee}$, one needs $m_1\approx -m_2$ and, to a good accuracy, $c^2_{12}\approx s^2_{12}$, in order to satisfy the bound on
$m_{ee}$.  This is exemplified by the following texture
\beq m_\nu =m
\left(
\begin{array}{ccc} 0& -1/\sqrt{2}& 1/\sqrt{2}\\ -1/\sqrt{2}& (1+\eta)/2& (1+\eta)/2\\ 1/\sqrt{2}& (1+\eta)/2& (1+\eta)/2
\end{array}
\right)~~~,
\label{deg3}
\eeq where $\eta\ll 1$, corresponding to an exact bimaximal mixing, $s_{13}=0$ and the eigenvalues are $m_1=m$, $m_2=-m$
and $m_3=(1+\eta) m$. This texture has been proposed in the context of a spontaneously broken SO(3) flavor symmetry and it
has been studied to analyze the stability of the degenerate spectrum against radiative corrections \cite{BRS,radcor}.  A more
realistic mass matrix can be obtained by adding small perturbations to $m_\nu$ in eq. (\ref{deg3}):
\beq m_\nu =m
\left(
\begin{array}{ccc}
\delta& -1/\sqrt{2}& (1-\epsilon)/\sqrt{2}\\ -1/\sqrt{2}& (1+\eta)/2& (1+\eta-\epsilon)/2\\ (1-\epsilon)/\sqrt{2}&
(1+\eta-\epsilon)/2&  (1+\eta-2\epsilon)/2
\end{array}
\right)~~~~~~~~~,
\label{deg4}
\eeq where $\epsilon$ parametrizes the leading flavor-dependent  radiative corrections (mainly induced by the $\tau$
Yukawa coupling) and $\delta$ controls $m_{ee}$. Consider first the case
$\delta\ll \epsilon$. 
 To first approximation $\theta_{12}$ remains maximal. We get $\Delta
m^2_{sun}\approx m^2 \epsilon^2/\eta$  and
\beq
\theta_{13}\approx 
\left(\frac{\Delta m^2_{sun}}{\Delta m^2_{atm}}
\right)^{{1}/{2}}~~~,~~~~~~~~~ m_{ee}\ll m~\left(\frac{\Delta m^2_{atm}~\Delta m^2_{sun}}{m^4}\right)^ {1/2}~~~.
\label{deg5}
\eeq If we instead assume $\delta\gg \epsilon$, we find $\Delta m^2_{sun}\approx  2 m^2 \delta$,
$\theta_{23}\approx\pi/4$, $\sin^2 2\theta_{12}\approx 1-\delta^2/4$. Also in this case the solar mixing angle remains unacceptably
close to $\pi/4$, unless further contributions to the mixing matrix are induced
from the diagonalization of the charged lepton sector. We get:
\beq
\theta_{13}\approx 0~~~,~~~~~~~~~~~~~ m_{ee}\approx \frac{\Delta m^2_{sun}}{2 m}~~~,
\label{deg6}
\eeq 
too small for detection if the average neutrino mass $m$ is around the eV scale. We see that with increasing $|m|$
more and more fine tuning is needed to reproduce the LA solution values of $\Delta m^2_{sun}$ and $\theta_{12}$. In
conclusion, even without invoking the WMAP limit, large
$\nu$ masses,
$|m| \sim (1\div 2)$ eV, are disfavoured by the limit on
$0\nu \beta
\beta$ decay and by the emerging of the LA solution with the solar angle definitely not maximal. From 
these considerations it is possible to derive the bound $|m|<0.9~ h$ eV (90\% C.L.) \cite{fsv}, 
where $h\approx 1$ parametrizes the uncertainties in the nuclear matrix elements.
Also, we have seen in sect. 5
that the attractive mechanism of baryogenesis through leptogenesis
appears to disfavour $|m|\gappeq 0.1$ eV,  at least in the simplest realizations. 
All together,
after WMAP and KamLAND, among degenerate models those with $|m|\lappeq (0.23\div 1)$ eV are favoured by converging evidence from
different points of view.

For $|m|$ not larger than the $0\nu \beta \beta$ bound one does not need a cancellation in $m_{ee}$ and $m_1$ and $m_2$
can be approximately equal in magnitude and phase. For example, in the limit $s_{13}=0$, the matrix
\beq m_\nu =m
\left(
\begin{array}{ccc}
1& 0& 0\\ 0&0& -1\\ 0&
-1&  0
\end{array}
\right)~~~,
\label{deg40}
\eeq 
corresponds to maximal $\theta_{23}$ (pseudo Dirac 23 sub-matrix) with $Diag[m_{\nu}]=m(1,1,-1)$. The angle
$\theta_{12}$ is unstable and a small perturbation can give any value to it. Note, however, that in this case the non
vanishing matrix elements must be of equal absolute value and not just of order 1. So either this is guaranteed by a
symmetry (as, for example, in ref. \cite{BHKR}) or the model is unnaturally fine-tuned.

As a different example (also with no cancellation between $m_1$ and $m_2$), a model, which is simple to describe but difficult
to derive in a natural way, is one
\cite{fd,bitri} where up quarks, down quarks and charged leptons have ``democratic'' mass matrices, with all entries equal (in first
approximation):
\beq  m_f = {\hat m}_f 
\left(
\matrix{ 1&1&1\cr  1&1&1\cr  1&1&1}
\right)+\delta m_f 
\label{dem}~~~,
\eeq  where ${\hat m}_f$ ($f=u,d,e$) are three overall mass parameters and $\delta m_f$ denote small perturbations. If we
neglect $\delta m_f$, the eigenvalues of $m_f$ are given by $(0,0,3~{\hat m}_f)$. The mass matrix $m_f$ is diagonalized by
a unitary matrix $U_f$ which is in part determined by the small term $\delta m_f$. If $\delta m_u\approx \delta m_d$, the
CKM matrix, given by $V_{CKM}=U_u^\dagger U_d$,  is nearly diagonal, due to a compensation between the large mixings
contained in 
$U_u$ and $U_d$. When the small terms $\delta m_f$  are diagonal and of the form $\delta m_f={\rm
Diag}(-\epsilon_f,\epsilon_f,\delta_f)$ with $\delta_f \gg \epsilon_f$, the matrices 
$U_f$ are approximately given by (note the analogy with the quark  model eigenvalues $\pi^0$, $\eta$ and $\eta'$):
\beq  U_f^\dagger\approx \left(
\matrix{ 1/\sqrt{2}&-1/\sqrt{2}&0\cr  1/\sqrt{6}&1/\sqrt{6}&-2/\sqrt{6}\cr  1/\sqrt{3}&1/\sqrt{3}&1/\sqrt{3}}
\right)~~~.
\label{ufri}
\eeq Note that, due to the degeneracy in the 1,2 sector in the unperturbed limit, any superposition of the first two rows would also be an eigenvector of zero mass. Thus the choice $(U_f^\dagger)_{13}=0$ is unjustified in the unperturbed limit and is only determined by a particular form of the perturbation. At the same time, the lightest quarks and charged leptons acquire a non-vanishing mass. The leading part of the mass
matrix in eq. (\ref{dem}) is invariant under a discrete
$S_{3L}\times S_{3R}$ permutation symmetry. The same requirement leads to the general neutrino mass matrix:
\beq m_{\nu} =m
\left[
\left(
\begin{array}{ccc} 1& 0& 0\\ 0& 1& 0\\ 0& 0& 1
\end{array}
\right) + r
\left(
\begin{array}{ccc} 1& 1& 1\\ 1& 1& 1\\ 1& 1& 1
\end{array}
\right)
\right] +\delta m_{\nu}~~~~~~~~~~~~,
\label{deg2}
\eeq where $\delta m_\nu$ is a small symmetry breaking term and the two independent invariants are allowed by the Majorana
nature of the light neutrinos. If $r$ vanishes the neutrinos are almost degenerate. In the presence of $\delta m_\nu$ the
permutation symmetry is broken and the degeneracy is removed.  If, for example, we choose $\delta m_{\nu}={\rm
Diag}(0,\epsilon,\eta)$ with $\epsilon<\eta\ll 1$ and $r\ll \epsilon$, the solar and the atmospheric oscillation
frequencies are determined by
$\epsilon$ and $\eta$, respectively. The problems
with this class of models are that the solar angle should be close to maximal (typically more than the atmospheric angle)
and that the neutrino spectrum and mixing angles are not determined by the symmetric limit (this applies, for example, to $s_{13}\sim 0$) but only by a
specific choice of the parameter $r$ and of the perturbations that cannot be easily justified on theoretical grounds.
Notice also that the simplest choice of parameters leads to 
$\sin^2 2\theta_{23}$ very close to 8/9, value which is now excluded at the 2
$\sigma$ level.

In this model the mixing angles are almost entirely due to the charged lepton sector. Recently the question of 
whether observed neutrino mixings can dominantly arise from the charged lepton sector in a natural way was studied 
in general in ref. \cite{noilast}. Of course, one can always choose an ad hoc basis where this is true: the point is 
to decide whether this formal choice can be naturally justified in the physical basis where the symmetries of the 
lagrangian are specified. The conclusion is that in presence of two large mixing angles $\theta_{12}$ and 
$\theta_{23}$ with the third angle $\theta_{13}$ being small, the construction of a natural model with dominance 
of $U_e$ is made much more difficult than in the case of only the atmospheric angle $\theta_{23}$ large.  
Exemples of natural models of this sort can be given \cite{AK,noilast,AK2,roma04} and
the stated difficulty is reflected in the relatively complicated symmetry structure required. 

In conclusion, the parameter space for degenerate models has recently become smaller because of the indications from 
WMAP (and also, too some extent, from leptogenesis) that tend to lower the maximum common mass allowed for light neutrinos. It is also rather difficult to
reproduce the observed pattern of frequencies and  mixing angles, in particular two large and one small mixing angle,
with the solar angle large but not maximal. Degenerate models that fit can only arise from a very special dynamics or a
non abelian flavour symmetry with suitable breakings.    

\subsection{Anarchy} 

Anarchical models \cite{anarchy} can be considered as particular cases of degenerate models with
$m^2\sim \Delta m^2_{atm}$. In this class of models mass degeneracy is replaced by the principle that all mass matrices are
structure-less in the neutrino sector, (including the LH charged fermions and possibly the RH neutrinos). For the LA solution the ratio of
the solar and atmospheric frequencies is not so small: $r={(\Delta m^2_{sun})}_{LA}/\Delta m^2_{atm}\sim 1/40$
and two out of three mixing angles are large. The key observation is that the see-saw mechanism tends to enhance the
ratio of eigenvalues: it is quadratic in $m_D$ so that a hierarchy factor $f$ in
$m_D$ becomes $f^2$ in $m_\nu$ and the presence of the Majorana matrix $M$ results in a further widening of the
distribution. Another squaring takes place in going from the masses to the oscillation frequencies which are quadratic. As
a result, a random generation of the $m_D$ and $M$ matrix elements leads to a distribution of $r$ that peaks around 0.1. At
the same time the distribution of $\sin^2{\theta_{ij}}$ is rather flat for all three mixing angles.
Clearly the smallness of $\theta_{13}$ is problematic for anarchy. This can be turned into the prediction that in
anarchical models
$\theta_{13}$ must be near the present bound (after all the value 0.2 for
$\sin{\theta_{13}}$ is not that smaller than the maximal value 0.707). In conclusion, if $\theta_{13}$ is near the present
bound, one can argue that the neutrino masses and mixings, interpreted by the see-saw mechanism, can just arise from
structure-less underlying Dirac and Majorana matrices.

\section{Inverted Hierarchy}

The inverted hierarchy configuration $|m_1|\sim |m_2| \gg |m_3|$ consists of two levels $m_1$ and
$m_2$ with small splitting $\Delta m_{12}^2~=~\Delta m_{sun}^2$ and a common mass given by
$\vert m_{1,2}^2\vert \sim \vert\Delta m_{atm}^2\vert\sim 2.6\cdot 10^{-3}$ eV$^2$. One particularly interesting example of
this sort \cite{invhier}, which leads to double maximal mixing, is obtained with the phase choice
$m_1=-m_2$ so that, approximately:
\beq  m_{diag}~=~{\rm Diag} (\sqrt{2} m,-\sqrt{2} m,0)~~~. 
\label{ih1}
\eeq The effective light neutrino mass matrix
\beq m_{\nu}~=~U^* m_{diag}U^\dagger\label{ih2}~~~,
\eeq which corresponds to the mixing matrix of double maximal mixing $c_{12}=s_{12}=1/\sqrt{2}$ and $s_{13}=u=0$ in eq.
(\ref{ufi1}).
\beq  U_{fi}= 
\left(\matrix{1/\sqrt{2}&1/\sqrt{2}&0 \cr -1/2& 1/2&1/\sqrt{2}\cr  1/2&-1/2&1/\sqrt{2}     } 
\right) ~~~,
\label{ufi2}
\eeq is given by:
\beq  m_{\nu}~=~m 
\left(\matrix{ 0&-1&1 \cr -1&0&0\cr 1&0&0     } 
\right) ~~~.
\label{ih3}
\eeq 
This texture for $m_{\nu}$ can be reproduced by imposing a U(1) flavour symmetry with charge $L_e-L_{\mu}-L_{\tau}$
starting either from $(H l)^T_i\lambda_{ij}(H l)_j/\Lambda$ or from RH neutrinos via the see-saw mechanism.  However the absolute values of the 12 and 13 terms would be in general different in this case. As a consequence, while the vanishing of $s_{13}$ and the maximal value of $\theta_{12}$ are still valid, the atmospheric angle deviates from the maximal 
value with $\tan{\theta_{23}}=x$ where $x$ is the absolute value of the ratio ${m_\nu}_{13}/{m_\nu}_{12}$.
We also note that the $1-2$ degeneracy remains stable under radiative corrections \cite{BRS,radcor}. 

The leading texture in (\ref{ih3}) can be perturbed by adding small terms:
\beq m_\nu =m
\left(
\begin{array}{ccc}
\delta& -1& 1\\ -1& \eta& \eta\\ 1& \eta& \eta
\end{array}
\right)~~~~~~~~~~~~~,
\label{invh1}
\eeq where $\delta$ and $\eta$ are small ($\ll 1$), real parameters defined up to  coefficients of order 1 that can differ
in the various matrix elements. One could also make the absolute values of the 12, 13 terms different by terms of order
$s_{13}\delta$. The perturbations leave
$\Delta m^2_{atm}$ and
$\theta_{23}$ unchanged, in first approximation. We obtain 
$\tan^2\theta_{12}\approx 1+\delta+\eta$ and 
$\Delta m^2_{sun}/\Delta m^2_{atm}\approx \eta+\delta$, where coefficients of order one have been neglected. Moreover
$\theta_{13}\approx\eta$. If $\eta\gg\delta$, we have
\beq
\theta_{13}\approx 
\frac{\Delta m^2_{sun}}{\Delta m^2_{atm}}~~~,~~~~~~~~~~~~~~m_{ee}\ll
\sqrt{\Delta m^2_{sun}} \left(\frac{\Delta m^2_{sun}}{\Delta m^2_{atm}}
\right)^{\frac{1}{2}} ~~~.
\label{ue3ih2}
\eeq In the other case, $\eta\ll\delta$ we obtain:
\beq
\theta_{13}\ll \frac{\Delta m^2_{sun}}{\Delta m^2_{atm}}~~~,~~~~~~~~~~~~~~ m_{ee}\approx\frac{1}{2}\sqrt{\Delta m^2_{sun}} 
\left(\frac{\Delta m^2_{sun}}{\Delta m^2_{atm}}\right)^{\frac{1}{2}}~~~.
\label{ue3ih1}
\eeq There is a well-known difficulty of this scenario to fit the  LA solution~\cite{invhier,bd,he}.  Indeed,
barring cancellation between the perturbations, in order to obtain a $\Delta m^2_{sun}$ close to the best fit LA
value, 
$\eta$ and $\delta$ should be smaller than about 0.1 and this keeps the value of $\sin^2 2\theta_{12}$ very close to 1, 
in disagreement with global fits of solar data~\cite{sunfit}. Notice that the required deviation of the solar 
mixing angle from the maximal value is of the order of the Cabibbo angle $\theta_C$ and indeed the empirical relation 
$\theta_{12}+\theta_C=\pi/4$ holds within the experimental errors \cite{raidal}. 
However, even starting from exact bimixing the
pattern of parameters needed to bring the solar angle down from the maximal value can be obtained in a natural way as an
effect of the charged lepton matrix diagonalization \cite{forinst}. 
This possibility, studied in detail in refs. \cite{noilast,frampton}, is not excluded but is strongly constrained by the observed smallness of $s_{13}$, as in general the amount of deviation from  maximal solar angle is typically of order $s_{13}$ (while the deviation from maximal atmospheric mixing are of second order).

With the phase choice $m_1=m_2$, i. e. for $Diag[m_{\nu}]=m(1,1,0)$, in the limit $s_{13}=0$, one obtains the matrix
\beq m_\nu =m
\left(
\begin{array}{ccc}
1& 0& 0\\ 0&1/2& -1/2\\ 0&
-1/2&  1/2
\end{array}
\right)~~~,
\label{deg60}
\eeq 
which corresponds to large $\theta_{23}$ with the solar angle unstable to small perturbations.
In this case fine tuning or a symmetry is necessary to fix the ratios of the matrix elements as indicated.

In conclusion, also for inverse hierarchy some special dynamics or symmetry is needed to reproduce in detail the observed features of the data. 

\section{Normal Hierarchy}

We now discuss the class of models which we consider the simplest approach to neutrino masses and mixings. In particular, in
this context one can formulate the most constrained framework which allows a comprehensive combined study of all fermion
masses in GUT's. We start by assuming three widely split $\nu$'s and the existence of a RH neutrino for each generation, as
required to complete a 16 dimensional representation of SO(10) for each generation. We then assume dominance of the see-saw
mechanism
$m_{\nu}=m_D^TM^{-1}m_D$. We know that the third-generation eigenvalue of the Dirac mass matrices of up and down quarks
and of charged leptons is systematically the largest one. It is natural (although not necessary) to imagine that this property
could also be true for the Dirac mass of $\nu$'s: $m_D^{diag}\sim {\rm Diag}(0,0,m_{D3})$. After see-saw we expect $m_{\nu}$
to be even more hierarchical being quadratic in
$m_D$ (barring fine-tuned compensations between $m_D$ and $M$). Note however that, for the LA solution: $r\sim 1/40$, so that the required amount of hierarchy,
$r=\Delta m^2_{atm} /\Delta m^2_{sun}=m^2_3/m^2_2$ is quite moderate. 

A possible difficulty
for hierarchical models is that one is used to expect that large splittings correspond to small mixings because normally
only close-by states are strongly mixed. In a 2 by 2 matrix context the requirement of large splitting and large mixings
leads to a condition of vanishing determinant and large off-diagonal elements. For example the matrix
\beq
\left(\matrix{ x^2&x\cr x&1    } 
\right) 
\label{md000}
\eeq has eigenvalues 0 and $1+x^2$ and for $x$ of O(1) the mixing is large. Thus in the limit of neglecting small mass
terms of order $m_{1,2}$ the demands of large atmospheric neutrino mixing and dominance of $m_3$ translate into the
condition that the 2 by 2 subdeterminant 23 of the 3 by 3 mixing matrix approximately vanishes. The problem is to show
that this vanishing can be arranged in a natural way without fine tuning. Once near maximal atmospheric neutrino mixing is
reproduced the solar neutrino mixing can be arranged to be either small or large without difficulty by implementing
suitable relations among the small mass terms. 
%\marginpar{aggiungere R-parity violation}

It is not difficult to imagine mechanisms that naturally lead to the approximate vanishing of the 23 sub-determinant. For
example in \cite{king,king2} it is assumed that one $\nu^c$ is particularly light and coupled to
$\mu$ and $\tau$. In a 2 by 2 simplified context if we have
\beq M\propto 
\left(\matrix{ \epsilon&0\cr 0&1    } 
\right)~~~,~~~~~~~ M^{-1}\approx\left(\matrix{ 1/\epsilon&0\cr 0&0    } 
\right)~~~,~~~~~~~m_D=\left(\matrix{a&b\cr c&d} 
\right)~~~,
\label{md0}
\eeq then for a generic $m_D$ we find
\beq m_{\nu}~=~m_D^TM^{-1}m_D\approx 
%\left(\matrix{ a&c\cr b&d   } 
%\right) \left(\matrix{ 1/\epsilon&0\cr 0&0    } 
%\right)\left(\matrix{ a&b\cr c&d   }\right)~=~
\frac{1}{\epsilon}\left(\matrix{ a^2&ab\cr ab&b^2   } 
\right)~~~.
\label{md1}
\eeq A different possibility that we find attractive is that, in the limit of neglecting terms of order
$m_{1,2}$ and, in the basis where charged leptons are diagonal, the Dirac matrix $m_D$, defined by $\nu^c m_D \nu$, takes
the approximate form, called ``lopsided''
\cite{lops1,lopsu1,lops2}:
\beq m_D\propto 
\left(\matrix{ 0&0&0\cr 0&0&0\cr 0&x&1    } 
\right)~~~~~. 
\label{md00}
\eeq This matrix has the property that for a generic Majorana matrix $M$ one finds:
\beq m_{\nu}=m^T_D M^{-1}m_D\propto 
\left(\matrix{ 0&0&0\cr 0&x^2&x\cr 0&x&1    } 
\right)~~~. 
\label{mn0}
\eeq The only condition on $M^{-1}$ is that the 33 entry is non zero.  However, when the approximately vanishing matrix
elements are replaced by small terms, one must also assume that no new O(1) terms are generated in $m_{\nu}$ by a
compensation between small terms in $m_D$ and large terms in
$M^{-1}$. 

It is important for the following discussion to observe that
$m_D$ given by eq. (\ref{md00}) under a change of basis transforms as $m_D\to V^{\dagger} m_D U$ where $V$ and $U$ rotate
the right and left fields respectively. It is easy to check that in order to make $m_D$ diagonal we need large left
mixings (i.e. large off diagonal terms in the matrix that rotates LH fields). Thus the question is how to reconcile large
LH mixings in the leptonic sector with the observed near diagonal form of $V_{CKM}$, the quark mixing matrix. Strictly
speaking, since $V_{CKM}=U^{\dagger}_u U_d$, the individual matrices $U_u$ and $U_d$ need not be near diagonal, but
$V_{CKM}$ does, while the analogue for leptons apparently cannot be near diagonal. However for quarks nothing forbids
that, in the basis where $m_u$ is diagonal, the $d$ quark matrix has large non diagonal terms that can be rotated away by
a pure RH rotation. We suggest that this is so and that in some way RH mixings for quarks correspond to LH mixings for
leptons.

In the context of (SUSY) SU(5) there is a very attractive hint of how this sort of  mechanism can be realized
\cite{lopsu5af,lopsu5}. In the
$\bar 5$ of SU(5) the $d^c$ singlet appears together with the lepton doublet $(\nu,e)$. The $(u,d)$ doublet and $e^c$
belong to the 10 and $\nu^c$ to the 1 and similarly for the other families. As a consequence, in the simplest model with
mass terms arising from only Higgs pentaplets, the Dirac matrix of down quarks is the transpose of the charged lepton
matrix:
$m_d=(m_l)^T$. Thus, indeed, a large mixing for RH down quarks corresponds to a large LH mixing for charged leptons.  At
leading order we may have the lopsided texture:
\beq  m_d=(m_l)^T=
\left(
\matrix{ 0&0&0\cr 0&0&1\cr 0&0&1}
\right) v_d~~~.
\eeq  In the same simplest approximation with  5 or $\bar 5$ Higgs, the up quark mass matrix is symmetric, so that left
and right mixing matrices are equal in this case. Then small mixings for up quarks and small LH mixings for down quarks
are sufficient to guarantee small $V_{CKM}$ mixing angles even for large $d$ quark RH mixings.  
%If these small mixings are neglected,
%we expect:
%\beq 
%m_u=
%\left(
%\matrix{ 0&0&0\cr 0&0&0\cr 0&0&1}
%\right) v_u~~~.
%\eeq 
%When the charged lepton matrix is diagonalized the large LH mixing of the charged
%leptons is transferred to the neutrinos. Note that in SU(5) we can diagonalize the $u$ mass matrix
%by a rotation of the fields in the 10, the Majorana matrix $M$ by a rotation of the 1 and the
%effective light neutrino matrix
%$m_\nu$ by a rotation of the $\bar 5$. In this basis the $d$ quark mass matrix fixes $V_{CKM}$ and
%the charged lepton mass matrix fixes neutrino mixings.  
It is well known that a model where the down and the charged
lepton matrices are exactly the transpose of one another cannot be exactly true because of the $e/d$ and
$\mu/s$ mass ratios. It is also known that one remedy to this problem is to add some Higgs component in the 45
representation of SU(5) \cite{jg}. But the symmetry under transposition can still be a good guideline if we are only
interested in the order of magnitude of the matrix entries and not in their exact values. In models with $U(1)_F$ flavour symmetry equal second and third generation charges for the $\bar 5$ could induce a lopsided form for both the charged leptons and the Dirac neutrino mass matrices. Otherwise, in the very crude model where the Higgs pentaplets come from a pure 10
representation of SO(10) one has
$m_D=m_u$ , i. e. the Dirac neutrino
mass matrix
$m_D$ is the same as the up quark mass matrix . For $m_D$ the dominance of the third family eigenvalue  as well as a near diagonal form could be an order of
magnitude remnant of this broken symmetry. 

To get a realistic mass matrix, we allow for deviations from the symmetric limit of (\ref{mn0}) with $x\sim o(1)$. For
instance, we can consider those models where the neutrino mass matrix elements are dominated, via the see-saw mechanism,
by the exchange of two right-handed neutrinos~\cite{king2}. Since the exchange of a single RH neutrino gives a successful
zeroth order texture, we are encouraged to continue along this line. Thus, we add a sub-dominant contribution of a second
RH neutrino, assuming that the third one gives a negligible contribution to the neutrino mass matrix, because it has much
smaller Yukawa couplings or is much heavier than the first two. The Lagrangian that describes this plausible subset of
see-saw models, written in the mass eigenstate basis of RH neutrinos and charged leptons, is
\beq {\cal L} = y_i \nu^c H l_i + y'_i {\nu^c}' H l_i +
\frac{M}{2} {\nu^c}^2 + \frac{M'}{2} {\nu^c}^{\prime 2}~~~,
\eeq leading to
\beq ({m_\nu})_{ij} \propto
\frac{y_i y_j}{M} +
\frac{y'_i y'_j}{M'}~~~,
\eeq where $i,j=\{e,\mu,\tau\}$. In particular, if $y_e\ll y_\mu\approx y_\tau$ and $y'_\mu\approx y'_\tau$, we obtain:
\beq m_\nu =m
\left(
\begin{array}{ccc}
\delta& \epsilon& \epsilon\\
\epsilon& 1+\eta& 1+\eta\\
\epsilon& 1+\eta& 1+\eta
\end{array}
\right)
\label{hier1}~~~~~~~~~~,
\eeq where coefficients of order one multiplying the small quantities 
$\delta$, $\epsilon$ and $\eta$ have been omitted. The 23 subdeterminant is generically of order $\eta$. The mass matrix in
(\ref{hier1}) does not describe the most general perturbation of the zeroth order texture  (\ref{mn0}). We have implicitly
assumed a symmetry between
$\nu_\mu$  and
$\nu_\tau$ which is preserved by the perturbations, at least at the level of the order of magnitudes. The perturbed
texture (\ref{hier1}) can also arise when the zeroes of the lopsided Dirac matrix in (\ref{md00}) are replaced by small
quantities.  It is possible to construct models along this line based on a spontaneously broken U(1)$_{\rm F}$ flavor
symmetry, where  
$\delta$, $\epsilon$ and $\eta$ are given by positive powers of one or more symmetry breaking parameters. Moreover, by
playing with the U(1)$_{\rm F}$ charges, we can adjust, to certain extent,  the relative hierarchy between $\eta$,
$\epsilon$ and 
$\delta$~\cite{king,king2,lopsu1,lops2,lopsu5af,lopsu5}, as we will see in section 8. The texture (\ref{hier1}) can also
be generated in SUSY models with $R$-parity violation \cite{rpv}. 

Let us come back to the mass matrix $m_\nu$ of eq.\ (\ref{hier1}). After a first rotation by an angle $\theta_{23}$ close
to 
$\pi/4$ and a second rotation with $\theta_{13}\approx \epsilon$, we get
\beq m_\nu
\approx m
\left(
\begin{array}{ccc}
\delta+\epsilon^2& \epsilon&0 \\
\epsilon& \eta& 0\\ 0& 0& 2
\end{array}
\right)
\label{hier2}~~~,
\eeq up to order one coefficients in the small entries. To obtain a large solar mixing angle, we need $|\eta-\delta|<
\epsilon$. In realistic models there is no reason for a cancellation  between independent perturbations and thus we assume
both
$\delta\le\epsilon$ and $\eta\le\epsilon$.  

Consider first the case $\delta\approx \epsilon$ and $\eta<\epsilon$. The solar mixing angle $\theta_{12}$ is large but
not maximal, as indicated by the LA solution. We also have $\Delta m^2_{atm}\approx 4 m^2$, $\Delta m^2_{sun}
\approx \Delta m^2_{atm} \epsilon^2$ and 
\beq m_{ee}\approx \sqrt{\Delta m^2_{sun}}~~~.
\label{bestnh}
\eeq

If $\eta\approx \epsilon$ and $\delta\ll\epsilon$,   we still have a large solar mixing angle and
$\Delta m^2_{sun}\approx \epsilon^2\Delta m^2_{atm}$, as before. However $m_{ee}$ will be much smaller than the estimate
in (\ref{bestnh}). This is the case of the models based on the above mentioned U(1)$_{\rm F}$ flavor
symmetry that, at least in its simplest realization, tends to predict $\delta\approx \epsilon^2$. In this class of models
we find 
\beq m_{ee}\approx \sqrt{\Delta m^2_{sun}}
\left(
\frac{\Delta m^2_{sun}}{\Delta m^2_{atm}}
\right)^{\frac{1}{2}}~~~, 
\label{nh2}
\eeq below the sensitivity of the next generation of planned  experiments. It is worth to mention that in both cases
discussed above, we have 
\beq
\theta_{13}\approx 
\left(\frac{\Delta m^2_{sun}}{\Delta m^2_{atm}}\right)^{\frac{1}{2}}~~~,
\label{ue3nh}
\eeq which might be close to the present experimental limit
if the oscillation frequency of the LA
solution for solar neutrinos is in the upper part of its allowed range.

If both $\delta$ and $\eta$ are much smaller than
$\epsilon$, the 12 block of $m_\nu$ has an approximate pseudo-Dirac structure and the angle $\theta_{12}$ becomes maximal.
This situation is typical of some models where leptons  have U(1)$_{\rm F}$ charges of both signs whereas the order
parameters of U(1)$_{\rm F}$ breaking have all charges of the same sign~\cite{lopsu5af}.  We have two eigenvalues
approximately given by $\pm m~\epsilon$. As an example, we consider the case where $\eta=0$ and 
$\delta\approx\epsilon^2$. We find $\sin^2 2\theta_{12}\approx 1-\epsilon^2/4$, $\Delta m^2_{sun}\approx m^2 \epsilon^3$
and
\beq
\theta_{13}\approx 
\left(\frac{\Delta m^2_{sun}}{\Delta m^2_{atm}}\right)^{\frac{1}{3}}~~~, ~~~~~~~~~~~ m_{ee}\approx \sqrt{\Delta m^2_{sun}}
\left(
\frac{\Delta m^2_{sun}}{\Delta m^2_{atm}}
\right)^{\frac{1}{6}}~~~. 
\label{nh3}
\eeq In order to recover the LA solution we would need a relatively large value of $\epsilon$. But this is in general
problematic because, on the one hand the presence of a large perturbation raises doubts about the consistency  of the
whole approach and, on the other hand, in existing models where all fermion sectors are related to each other, $\epsilon$
is never larger than the Cabibbo angle. 

Summarising, within normal hierachical models there is enough flexibility to reproduce in a natural way the experimental
frequencies and mixing angles. In particular the lopsided matrix solution of the large atmospheric mixing, inspired by SU(5),
where the large atmospheric mixing arises from the charged lepton sector, can be extended rather naturally to also account for the solar
sector and for the small $\theta_{13}$ mixing angle. 

\subsection{Semi-anarchy}

We have seen that anarchy is the absence of structure in the neutrino sector. Here we consider an attenuation of anarchy where
the absence of structure is limited to the 23 sector. The typical texture is in this case: 
\beq m_\nu
\approx m
\left(
\begin{array}{ccc}
\delta& \epsilon&\epsilon \\
\epsilon& 1&1\\ \epsilon& 1& 1
\end{array}
\right)
\label{s-an}~~~,
\eeq
where $\delta$ and $\epsilon$ are small and by 1 we mean entries of $O(1)$ and also the 23 determinant is of $O(1)$. 
We see
that this texture is similar to eq. (\ref{hier1}) when $\eta\sim O(1)$. Clearly, in general we would
expect two mass eigenvalues of order 1, in units of $m$, and one small, of order $\delta$ or $\epsilon^2$. 
This pattern does not fit the observed solar
and atmospheric observed frequencies. However, given that the ratio
$r={(\Delta m^2_{sun})}_{LA}/\Delta m^2_{atm}\sim 1/40$ is not too small, we can assume that the small value of $r$ is
generated accidentally, as for anarchy. We see that, if we proceed with the same change of basis as from eq. (\ref{hier1})
to eq. (\ref{hier2}), it is sufficient that by chance $\eta\sim \delta+\epsilon^2$ in order to obtain the correct value of $r$
with large $\theta_{23}$ and $\theta_{12}$ and small $\theta_{13}\sim \epsilon$. The natural smallness of $\theta_{13}$
is the main advantage over anarchy. We will come back to this class of models in a following section.   

\section{Grand Unified Models of Fermion Masses}

We have seen that the smallness of neutrino masses interpreted via the see-saw mechanism directly leads to a scale 
$\Lambda$ 
for L non-conservation which is remarkably close to $M_{GUT}$. Thus neutrino masses and mixings should find a
natural context in a GUT treatment of all fermion masses. The hierarchical pattern of quark and lepton masses, within
a generation and across generations, requires some dynamical suppression mechanism that acts differently on the
various particles. 
This hierarchy can be generated by a number of operators of different dimensions suppressed by inverse
powers of the cut-off $\Lambda_c$ of the theory. In some realizations, the different powers of $1/\Lambda_c$
correspond to different orders in some symmetry breaking parameter $v_f$ arising from the spontaneous 
breaking of a flavour symmetry.
In the next subsections we describe some simplest models based
on SU(5) $\times$ U(1)$_{\rm F}$ and on SO(10) which illustrate these possibilities \cite{nonab}.
It is notoriously difficult to turn these models into fully realistic theories, due to
well-known problems such as the doublet-triplet splitting, the proton lifetime, the gauge coupling unification 
beyond leading order and the wrong mass relations for charged fermions of the first two  generations.
Some of these problems can be solved by adopting the elegant idea of GUT's in extra dimensions \cite{EDGUTs}. Here we adopt the GUT framework simply as a convenient testing ground for different neutrino mass scenarios.
 
\subsection{Models Based on Horizontal Abelian Charges}
 
We discuss here some explicit examples of grand unified models in the framework of a unified SUSY
SU(5) theory with an additional 
U(1)$_{\rm F}$ flavour symmetry. The SU(5) generators  act ``vertically'' inside one generation, while the
U(1)$_{\rm F}$ charges are different ``horizontally'' from one generation to the other. If, for a
given interaction vertex, the
U(1)$_{\rm F}$ charges do not add to zero, the vertex is forbidden in the symmetric limit. But the symmetry is spontaneously broken by the VEV's
$v_f$ of a number of ``flavon'' fields with non-vanishing charge. Then a forbidden coupling is rescued but is
suppressed by powers of the small parameters $v_f/\Lambda_c$ with the exponents larger for larger charge 
mismatch \cite{u1}. We expect $M_{GUT} \lappeq v_f \lappeq \Lambda_c \lappeq M_{Pl}$. 
Here we discuss some aspects of the description of fermion masses in this framework. 

In these models the known generations of quarks and leptons are contained in triplets
$\Psi^{10}_i$ and
$\Psi^{\bar 5}_i$, $(i=1,2,3)$ corresponding to the 3 generations, transforming as $10$ and ${\bar 5}$ of SU(5),
respectively. Three more
SU(5) singlets
$\Psi^1_i$ describe the RH neutrinos. In SUSY models we have two Higgs multiplets $H_u$ and $H_d$, which transform as 5
and $\bar 5$ in the minimal model. The two Higgs multiplets may have the same or different charges.
In all the models that we discuss the large atmospheric mixing angle
is described by assigning equal flavour charge to muon and tau neutrinos and their weak SU(2) partners (all belonging
to the 
${\bar 5}\equiv(l,d^c)$ representation of SU(5)). Instead, the solar neutrino oscillations can be obtained with
different, inequivalent charge assignments.
There are many variants of these models: fermion charges can all be non-negative with only negatively charged
flavons, or there can be fermion charges of different signs with either flavons of both charges or only flavons of
one charge. We can have that only the top quark mass is allowed in the symmetric limit, or that also other third
generation fermion masses are allowed. The Higgs charges can be equal, in particular both vanishing or can be
different. We can arrange that all the structure is in charged fermion masses while neutrinos are anarchical.  

\subsubsection{F(fermions)$\ge$ 0}

Consider, for example, a simple model with all charges of matter fields being non-negative and containing one
single flavon ${\bar\theta}$ of charge F$=-1$. For a maximum of simplicity we also assume that all the  third generation masses
are directly allowed in the symmetric limit. This is realized by taking vanishing charges for the Higgses and for
the third generation components $\Psi^{10}_3$, $\Psi^{\bar 5}_3$ and $\Psi^1_3$.  
If we define F$(\Psi^R_i)\equiv q^R_i$ $(R=10,{\bar 5},1;~i=1,2,3)$,
then the generic mass matrix $m$ has the form
\beq 
m~=~
\left(
\matrix{ y_{11}\lambda^{q^R_1+q^{R'}_1}&y_{12}\lambda^{q^R_1+q^{R'}_2}&y_{13}\lambda^{q^R_1+q^{R'}_3}\cr
y_{21}\lambda^{q^R_2+q^{R'}_1}&y_{22}\lambda^{q^R_2+q^{R'}_2}&y_{23}\lambda^{q^R_2+q^{R'}_3}\cr
y_{31}\lambda^{q^R_3+q^{R'}_1}&y_{32}\lambda^{q^R_3+q^{R'}_2}&y_{33}\lambda^{q^R_3+q^{R'}_3}}
\right) v~~~,
\label{m1}
\eeq 
where all the $y_{ij}$ are dimensionless complex coefficients of order one and
$m_u$, $m_d=m_l^T$, $m_D$ and $M$ arise by choosing $(R,R')=(10,10)$, $(\bar{5},10)$,  
$(1,\bar{5})$ and $(1,1)$, respectively. 
We have $\lambda\equiv\langle{\bar \theta}\rangle/\Lambda_c$ and the 
quantity $v$ represents the appropriate VEV or mass parameter.
The models with all non-negative charges and one single flavon have particularly simple factorization properties.
For instance in the see-saw expression for $m_{\nu}=m_D^T M^{-1} m_D$ the dependence
on the $q^{1}_i$ charges drops out and only that from $q^{\bar5}_i$ remains.
In addition, for the neutrino mixing matrix $U_{ij}$, which is determined by $m_\nu$ in
the basis where the charged leptons are diagonal, one can prove that 
$U_{ij}\approx \lambda^{|q^{\bar 5}_i-q^{\bar 5}_j|}$, in terms of the differences of the
$\bar5$ charges, when terms that are down by powers of
the small parameter $\lambda$ are neglected. Similarly the CKM matrix elements are approximately 
determined by only the 10 charges \cite{u1}: $V^{CKM}_{ij}\approx\lambda^{|q^{10}_i-q^{10}_j|}$.
If the symmetry breaking parameter $\lambda$ is numerically close to the Cabibbo angle,
we can choose:
\beq
(q^{10}_1,q^{10}_2,q^{10}_3)=(3,2,0)~~~,
\label{cha10}
\eeq     
thus reproducing $V_{us}\sim\lambda$, $V_{cb}\sim\lambda^2$ and
$V_{ub}\sim\lambda^3$. The same $q^{10}_i$ charges also fix $m_u:m_c:m_t\sim \lambda^6:\lambda^4:1$. The
experimental value of $m_u$ (the relevant mass values are those at the GUT scale: $m=m(M_{GUT})$ \cite{koide}) 
would rather prefer $q^{10}_1=4$. Taking into account this indication and the presence of the unknown coefficients 
$y_{ij}\sim O(1)$ it is difficult to decide between $q^{10}_1=3$ or $4$ and both are acceptable. 
Of course the charges $(q^{10}_1,q^{10}_2,q^{10}_3)=(2,1,0)$ would represent an equally good
choice, provided we appropriately rescale the expansion parameter $\lambda$.
Turning to the $\bar 5$ charges, if we take \cite{lopsu1,lops2,lopsu5af,u1again,chargesnu}
\beq
(q^{\bar 5}_1,q^{\bar 5}_2,q^{\bar 5}_3)=(b,0,0)~~~~~~~~~~~~b\ge0~~~, 
\label{cha5}
\eeq
together with eq. (\ref{cha10}) we get the patterns $m_d:m_s:m_b\sim m_e:m_\mu:m_\tau \sim 
\lambda^{3+b}:\lambda^2:1$. Moreover,
the 22, 23, 32, 33 entries of the effective light neutrino mass matrix $m_\nu$ are all O(1), thus accommodating 
the nearly maximal value of $s_{23}$. The small non diagonal terms of the
charged lepton mass matrix cannot change this. 
We obtain, where arbitrary O(1) coefficients are omitted:   
\beq 
m_\nu=
\left(
\matrix{
\lambda^{2b}&\lambda^b&\lambda^b\cr
\lambda^b&1&1\cr
\lambda^b&1&1}\right){v_u^2\over \Lambda}~~~~~~~({\tt A,SA})~~~, 
\label{mlep}
\eeq 
where $v_u$ is the VEVs of the Higgs doublet giving mass to the up quarks and all the entries are specified 
up to order one coefficients. If we take $v_u\sim 250~{\rm GeV}$, the mass scale ${\Lambda}$ of
the heavy Majorana neutrinos turns out to be close to the unification scale, 
${\Lambda}\sim 10^{15}~{\rm GeV}$.

If $b$ vanishes, then the light neutrino mass matrix will be structure-less and we recover the anarchical (A)
picture of neutrinos discussed in section 6.1. In a large sample of anarchical models, generated with 
random coefficients, the resulting neutrino mass spectrum can exhibit either normal or inverse hierarchy. 
For down quarks and charged leptons we obtain a weakened hierarchy, essentially the square root than that of up quarks.

If $b$ is positive, then the light neutrino mass matrix will be structure-less only in the (2,3) sub-sector and we
get semi-anarchical (SA) models, introduced in section 7.2. 
In this case, the neutrino mass spectrum has normal hierarchy. However,
unless the (2,3) sub-determinant is accidentally suppressed, atmospheric and solar oscillation frequencies are 
expected to be of the same order and, in addition, the preferred solar mixing angle is small. Nevertheless, such a
suppression can occur in a fraction of semi-anarchical models generated with random, order one coefficients. The
real advantage over the fully anarchical scheme is represented by the suppression in $U_{e3}$. 

Note that in all previous cases we could add a constant to $q^{\bar 5}_i$, for example by taking 
$(q^{\bar5}_1,q^{\bar5}_2,q^{\bar5}_3)
=  (2+b,2,2)$. This would only have the consequence to leave the top quark as the only unsuppressed mass and to
decrease the resulting value of $\tan{\beta}=v_u/v_d$ down to $\lambda^2 m_t/m_b$. A constant shift of the charges $q^1_i$ might also provide a suppression
of the leading $\nu^c$ mass eigenvalue, from $\Lambda_c$ down to the 
appropriate scale $\Lambda$. One
can also consider models where the 5 and $\bar 5$ Higgs charges are different, as in the ``realistic'' SU(5) model of
ref. \cite{afm2}. Also in these models the top mass could be the only one to be non-vanishing in the symmetric limit and the
value of $\tan{\beta}$ can be adjusted.

\subsubsection{F(fermions) and F(flavons) of both signs}

Models with naturally large 23 splittings are obtained if we allow negative charges and, at the same time, either
introduce flavons of opposite charges or stipulate that matrix elements with overall negative charge are put to
zero. For example, we can assign to the fermion fields the set of
F charges given by:
\bea
(q^{10}_1,q^{10}_2,q^{10}_3) &= & (3,2,0) \nn\\
(q^{\bar 5}_1,q^{\bar 5}_2,q^{\bar 5}_3) &= & (b,0,0)~~~~~~~~~~~~b\ge 2a>0\nn\\
(q^{1}_1,q^{1}_2,q^{1}_3) &= & (a,-a,0) ~~~.\label{cha1}
\eea   
We consider the Yukawa coupling allowed by U(1)$_{\rm F}$-neutral  Higgs multiplets
in the $5$ and ${\bar 5}$ SU(5) representations and by a pair $\theta$ and
${\bar\theta}$ of SU(5) singlets with F$=1$ and F$=-1$, respectively. 
If $b=2$ or 3, the up, down and charged lepton sectors are
not essentially different than in the SA case. Also in this case the O(1) off-diagonal entry of $m_l$, typical
of lopsided models, gives rise to a large LH  mixing in the 23 block which corresponds to a large
RH mixing in the
$d$ mass matrix. In the neutrino sector, 
after diagonalization of the charged lepton sector and after
integrating out the heavy RH neutrinos we obtain the following neutrino mass matrix in the low-energy
effective theory:
\beq
m_\nu=
\left(
\matrix{
\lambda^{2 b}&\lambda^b&\lambda^b\cr
\lambda^b&1+\lambda^a{\lambda'}^a&1+\lambda^a{\lambda'}^a\cr
\lambda^b&1+\lambda^a{\lambda'}^a&1+\lambda^a{\lambda'}^a}\right)
{v_u^2\over {\Lambda}}~~~~~~~({\tt H}),
\label{mnu}
\eeq
where $\lambda'$ is given by $\langle\theta\rangle/\Lambda_c$ and ${\Lambda}$ as before denotes the large mass 
scale associated to the
RH neutrinos: ${\Lambda}\gg v_{u,d}$. 
The O(1) elements in the 23 block are produced by combining the
large  LH mixing induced by the charged lepton sector and the large LH mixing in $m_D$. A crucial
property of $m_\nu$ is that, as a result of the see-saw mechanism and of the specific U(1)$_{\rm F}$ 
charge assignment, the determinant of the 23 block is automatically of $O(\lambda^a{\lambda'}^a)$ 
(for this the presence of negative
charge values, leading to the presence of both $\lambda$ and
$\lambda'$ is essential \cite{lops2,lopsu5af}). The neutrino mass matrix of eq. 
(\ref{mnu}) is a particular case of the more general pattern 
presented in eq. (\ref{hier1}), for $\delta\approx\lambda^{2b}$,
$\epsilon\approx\lambda^b$ and $\eta\approx\lambda^a{\lambda'}^a$. If we take
$\lambda\approx\lambda'$, it is easy to verify that the eigenvalues of $m_\nu$ satisfy  the relations:
\beq m_1:m_2:m_3  = \lambda^{2(b-a)}:\lambda^{2a}:1~~.
\eeq 
The atmospheric neutrino oscillations require 
$m_3^2\sim 10^{-3}~{\rm eV}^2$. The squared mass difference between the lightest states is  of
$O(\lambda^{4a})~m_3^2$, not far from the LA solution to the solar neutrino problem if we choose $a=1$. In general $U_{e3}$ is
non-vanishing, of $O(\lambda^b)$. Finally, beyond the large mixing in the 23 sector,
$m_\nu$  provides a mixing angle $\theta_{12} \sim \lambda^{b-2a}$ 
in the 12 sector. 
When $b=2 a$, as for instance in the case 
$b=2$ and $a=1$, the  LA solution can be reproduced and 
the resulting neutrino spectrum is hierarchical (H).

Alternatively, an inversely hierarchical (IH) spectrum can be obtained by choosing:
\bea
(q^{10}_1,q^{10}_2,q^{10}_3) &= & (3,2,0) \nn\\
(q^{\bar 5}_1,q^{\bar 5}_2,q^{\bar 5}_3) &= & (1,-1,-1)~~~~~~~~~~~~\nn\\
(q^{1}_1,q^{1}_2,q^{1}_3) &= & (-1,1,0) \nn\\
(q_{H_u},q_{H_d})&=&(0,1)~~~.\label{cha2}
\eea   
Due to the non-vanishing charge of the $H_d$ Higgs doublet, in the charged lepton sector
we recover the same pattern previously discussed.
The light neutrino mass matrix is given by:
\beq
m_\nu=
\left(
\begin{array}{ccc}
\lambda^2 & 1 & 1\\
1 & \lambda'^2 & \lambda'^2\\
1 & \lambda'^2 & \lambda'^2
\end{array}
\right)~~~~~~~~~~({\tt IH})~~~.
\label{mnuih}
\eeq
The ratio between the solar and atmospheric
oscillation frequencies is not directly related to the sub-determinant 
of the block 23, in this case.

\vspace{0.1cm}
\begin{table}[!h]
\caption{Models and their flavour charges. \label{updt_tab1}}
\vspace{0.4cm}
\begin{center}
\begin{tabular}{|c|c|c|c|c|}
\hline 
& & & & \\ 
{\tt Model}& ${{\Psi_{10}}}$ & ${\Psi_{\bar 5}}$ & ${{\Psi_1}}$ & ${(H_u,H_d)}$ \\ 
& & & & \\
\hline
\hline
 & & & & \\ 
{\tt Anarchical ($A$)}& (3,2,0)& (0,0,0) & (0,0,0) & (0,0)\\ 
& & & & \\
\hline
 & & & & \\ 
{\tt Semi-Anarchical ($SA$)}& (2,1,0) & (1,0,0) & (2,1,0) & (0,0) \\ 
& & & & \\
\hline
\hline 
& & & & \\
{\tt Hierarchical ($H_{I}$)}& (6,4,0)& (2,0,0) & (1,-1,0) & (0,0)\\ 
& & & & \\
\hline
& & & & \\
{\tt Hierarchical ($H_{II}$)}& (5,3,0)& (2,0,0) & (1,-1,0) & (0,0)\\ 
& & & & \\
\hline
& & & & \\ 
{\tt Inversely Hierarchical ($IH_{I}$)}& (3,2,0) & (1,-1,-1)& (-1,+1,0)& (0,+1) \\
& & & & \\
\hline& & & & \\ 
{\tt Inversely Hierarchical ($IH_{II}$)}& (6,4,0) & (1,-1,-1)& (-1,+1,0)& (0,+1) \\
& & & & \\
\hline
\end{tabular}
\end{center}
\end{table}

A representative set of models is listed in table 2. Note that in some cases the charges for $\Psi_{10}$ have been changed from $(3,2,0)$  (our reference values in eqs. (\ref{cha10}), (\ref{cha1}), and (\ref{cha2})) to $(6,4,0)$ or $(5,3,0)$. These values are a posteriori better suited to reproduce the moderate level of hierarchy implied by the present neutrino oscillation data. Since the neutrino mixing parameters are completely independent on
the 10 charges, this change is only important for a
better fit to quark and charged lepton masses and mixings once a rather large value of $\lambda$ is derived from the neutrino data.  
The hierarchical and the inversely hierarchical models may come into  several varieties depending on the number and
the charge of the flavour symmetry breaking (FSB) parameters. Above we have considered the case
of two (II)  oppositely charged flavons with symmetry breaking
parameters $\lambda$ and $\lambda'$. It may be noticed that the presence of two multiplets $\theta$ and
${\bar \theta}$ with opposite F charges could hardly be reconciled, without adding extra structure to the model,
with a large common VEV for these fields, due to possible analytic terms of the kind $(\theta {\bar \theta})^n$ in
the superpotential. Therefore it is instructive to explore the consequences of allowing only the negatively
charged ${\bar \theta}$ field in the theory, case I.
In case I, it is impossible to compensate negative F charges in the Yukawa
couplings and the corresponding entries in  the neutrino mass matrices vanish. Eventually these zeroes are filled by
small contributions, arising, for instance, from the diagonalization of the charged lepton sector or from the
transformations needed to make the kinetic terms canonical. 

Another important ingredient is represented by the see-saw mechanism \cite{seesaw}. Hierarchical models and
semi-anarchical  models have similar charges in the $(10,{\bar 5})$ sectors and, in the absence of the see-saw
mechanism, they would give rise to similar results. Even when the results are expected to be independent from the
charges of the RH neutrinos, as it is the case for the anarchical and semi-anarchical models, the see-saw
mechanism can induce some sizeable effect in a statistical  analysis. For this reason, for each type of model, but
the normal-hierarchical ones (the mechanism for the 23 sub-determinant suppression is in fact based on the see-saw
mechanism), it is interesting to study the case where RH neutrinos are present and the see-saw contribution 
is the dominant one (SS) and the case where they are absent and the mass matrix is saturated by the 
non-renormalizable contribution (NOSS). 

With this classification in mind, we can distinguish the following type of models, all
supported by specific choices of U(1) charges:
${\rm A_{SS}}$, ${\rm A_{NOSS}}$, ${\rm SA_{SS}}$, ${\rm SA_{NOSS}}$,
${\rm H_{(SS,I)}}$, ${\rm H_{(SS,II)}}$, ${\rm IH_{(SS,I)}}$, 
${\rm IH_{(SS,II)}}$, ${\rm IH_{(NOSS,I)}}$ and ${\rm IH_{(NOSS,II)}}$.

\begin{figure}[!h]
\centerline{   \psfig{file=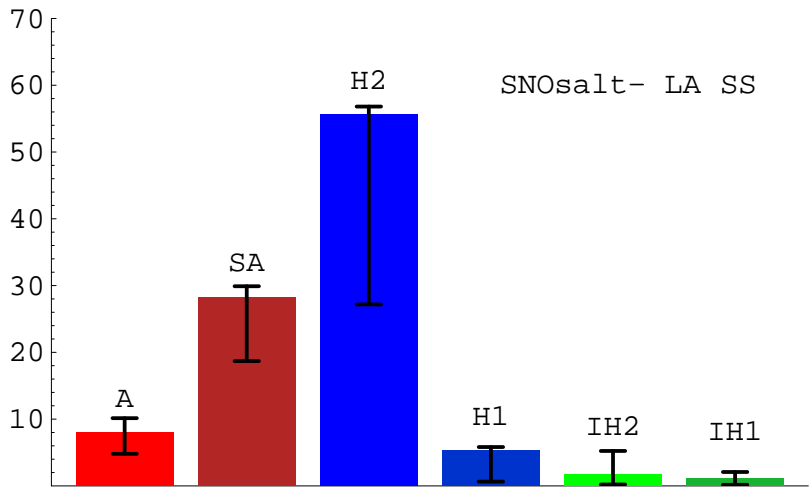,width=0.7 \textwidth}  }
\caption{Relative success rates for the LA solution, with see-saw. 
The sum of the rates has been normalized to 100. The results correspond to the 
default choice ${\cal I}=[0.5,2]$, and to the following values 
of $\lambda=\lambda'$: $0.2$, $0.25$, $0.35$, $0.45$, $0.45$, $0.25$ 
for the models ${\rm A_{SS}}$, ${\rm SA_{SS}}$, ${\rm H_{(SS,II)}}$, 
${\rm H_{(SS,I)}}$, ${\rm IH_{(SS,II)}}$ and ${\rm IH_{(SS,I)}}$, 
respectively. The error bars represent
the linear sum of the systematic error due to the
choice of ${\cal I}$ and the statistical error (see text).}
\label{updt_barlass}
\end{figure}

\begin{figure}[!h]
\vskip .5 cm
\centerline{  \psfig{file=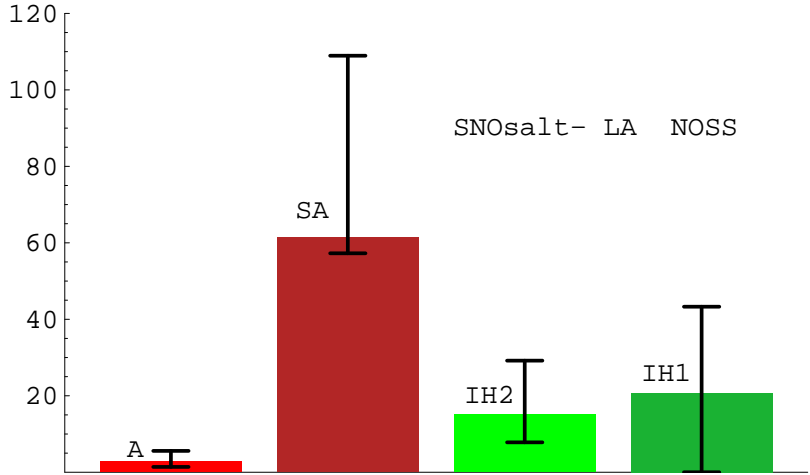,width=0.7 \textwidth}   }
\caption{Relative success rates for the LA solution, without see-saw. 
The sum of the rates has been normalized to 100. The results correspond to the 
default choice ${\cal I}=[0.5,2]$, and to the following values 
of $\lambda=\lambda'$: $0.2$, $0.2$, $0.25$, $0.25$
for the models ${\rm A_{NOSS}}$, ${\rm SA_{NOSS}}$, ${\rm IH_{(NOSS,II)}}$, 
and ${\rm IH_{(NOSS,I)}}$, respectively (in our notation there are no ${\rm H_{(NOSS,I)}}$, 
${\rm H_{(NOSS,II)}}$ models). 
The error bars represent the linear sum of the systematic error due to the
choice of ${\cal I}$ and the statistical error (see text).}
\label{updt_barlanoss}
\end{figure}

It is interesting to quantify the ability of each model in reproducing the observed oscillation
parameters. For anarchy, it has been observed that random generated, order-one 
entries of the neutrino mass matrices (in appropriate units), correctly fit the experimental data
with a success rate of few percent. It is natural to extend this analysis to include also
the other models based on SU(5) $\times$ U(1) \cite{afm03}, which have mass matrix elements defined up to 
order-one dimensionless coefficients $y_{ij}$ (see eq. \ref{m1}). For each model, successful 
points in parameter space are selected by asking that the four observable quantities $O_1=r
\equiv \Delta m^2_{12}/\vert\Delta m^2_{23}\vert$,
$O_2=\tan^2\theta_{12}$, $O_3=\vert U_{e3}\vert\equiv\vert\sin\theta_{13}\vert$ and 
$O_4=\tan^2\theta_{23}$ fall in the approximately 3$\sigma$ allowed ranges \cite{sunfit,atmfit}: 
\beq
\begin{array}{l}
 0.018 < r < 0.053\\
\vert U_{e3} \vert < 0.23\\ 0.30 <\tan^2\theta_{12}<0.64\\ 0.45 <\tan^2\theta_{23}<2.57
\end{array}
\label{lanew}
\eeq
The coefficients $y_{ij}$ of the neutrino sector are random complex numbers with absolute values
and phases uniformly distributed in intervals ${\cal I}=[0.5,2]$ and $[0,2\pi]$ respectively.
The dependence of the results on these choices can be estimated by varying
${\cal I}$.
For each model an optimization procedure selects the value of the flavour symmetry breaking
parameter $\lambda=\lambda'$ that maximizes the success rate.
The success rates are displayed in figs. \ref{updt_barlanoss}  and \ref{updt_barlass}, separately for the SS and NOSS
cases. The two sets of models have been individually normalized to give a total rate 100. From the 
histograms in figs. \ref{updt_barlass} and  \ref{updt_barlanoss}  we see that normal hierarchy models  
are neatly
preferred over anarchy and inverse hierarchy in the context of these SU(5)$\times$U(1) models. In particular, in the SS
case, the ${\rm H_{II}}$ models with normal hierarchy, two oppositely charged flavons
and suppressed 23 sub-determinant are clearly preferred. Models of the type ${\rm H_{I}}$ are disfavoured. For the relatively large values of the expansion parameter
required to fit $r$, they tend to predict too large $\vert U_{e3} \vert$ and $\tan^2\theta_{12}>1$.
We recall that for the chosen charge values the ${\rm H_{II}}$ model is of the lopsided type.
In the NOSS case the see-saw suppression of the 23 determinant is clearly not operative and all normal 
hierarchy models coincide with SA.

\vskip 0.3cm

An interesting question is whether the disfavouring of IH models that we find in our SU(5)$\times$U(1) framework can be
extended to a more general context. In the limit of vanishing $\lambda$ and $\lambda'$ the IH texture (see eq. (\ref{mnuih})) 
becomes close to that of
bimaximal mixing and $\theta_{13}=0$ (actually with $r=0$). In our U(1) models $r\approx
\vert U_{e3}\vert \approx\vert\tan^2\theta_{12}-1\vert \approx O(\lambda^2)$ (for $\lambda=\lambda'$). In particular the charged lepton mixings
cannot displace too much $\theta_{12}$ from its maximal value because the small value of the electron mass forces a
sufficiently large value of the relevant charges, which in turn implies that the charged lepton mixing correction to
$\theta_{12}$ is small. We have already mentioned that corrections from the charged lepton sector can in principle bring the predictions of a neutrino matrix of the bimixing type in agreement with the data and that the smallness of $s_{13}$ induces strong constraints. In the particular setup of U(1)$_{\rm F}$ models we have seen that charged lepton corrections are too small to make the solar angle sufficiently different from maximal.

Leptonic mixing in SUSY GUTs with RH neutrinos are potential sources
of lepton flavour violating (LFV) processes beyond neutrino oscillations 
\cite{LFV}.
The observable effects are difficult to estimate since they are sensitive to
both the details of the SUSY breaking mechanism and to the low-energy
superparticle spectrum. In models with gravity mediated SUSY breaking
and universal boundary conditions for the soft breaking terms at the cut-off
scale $\Lambda$, running effects give rise to off-diagonal slepton masses
$\delta {m_{ij}^2}^{LL}$ proportional, at leading order, to 
$C_{ij}=(y_\nu^\dagger \log(\Lambda/M) y_\nu)_{ij}$.
Current bounds on LFV transitions $l_i\to l_j\gamma$ translate into
an upper bound on the combination $\vert C_{ij}\vert$, 
depending on $\tan\beta$ and soft mass parameters. If, for instance,
$\tan\beta=10$, the present experimental bound on $BR(\mu\to e\gamma)$
already excludes $C_{21}>1$ for the most plausible values of slepton
and gaugino masses. In the models considered in this section
$C_{21}$ is dominated by the couplings to the RH neutrino of the third 
generation and, assuming $(\Lambda/M_3)\approx O(100)$, we roughtly 
expect $\vert y_{32}^* y_{31}\vert<0.2$. This constraint is most
easily respected by models with inverse hierarchy. For semianarchy and normal
hierarchy the constraint is almost saturated, whereas anarchy
tends to violate it. Future improvements in the experimental sensitivity
could lead to a significant selection of the models.

In conclusion, models based on SU(5) $\times$ U(1)$_{\rm F}$ are clearly toy models that can only aim at a semiquantitative
description of fermion masses. In fact only the order of magnitude of each matrix entry can be specified. However
it is rather impressive that a reasonable description of fermion masses, now also including neutrino masses and
mixings, can be obtained in this simple context, which is suggestive of a deeper relation between gauge and flavour
quantum numbers. There are 12 mass eigenvalues and 6 mixing angles that are specified, modulo coefficients of
order 1, in terms of a bunch of integer numbers (from half a dozen to a dozen), the charges, plus 1 or more scale
parameters. Moreover all possible type of mass hierarchies can be reproduced within this framework.
In a statistically based comparison, the range of $r$ and the small upper
limit on
$U_{e3}$  are sufficiently constraining to make anarchy neatly disfavoured with respect to models
with built-in hierarchy. If only neutrinos are considered, one might counterargue that hierarchical models have
at least one more parameter than anarchy, in our case the parameter $\lambda$. However, if one looks at quarks and
leptons together, as in the GUT models that we consider, then the same parameter that plays the role of an order parameter
for the CKM matrix, for example, the Cabibbo angle, can be successfully used to reproduce also the hierarchy
implied by the present neutrino data.

\subsection{GUT Models based on SO(10)}

The fermion sector of SO(10) grand unified theories has remarkable properties.
It is automatically anomaly-free, independently from the representation content. The RH neutrino
provides the completion of a SM family into a 16 representation, thus
offering a natural ground for the see-saw mechanism and baryogenesis through leptogenesis.
Moreover, the scale of lepton number violation, determined by the gauge symmetry breaking pattern, 
can be smaller than the cut-off $\Lambda_c$ of the theory, and closer to the grand unification scale itself 
\cite{buc}, 
as suggested by the mass scale associated to atmospheric neutrino oscillations.

In their simplest realizations SO(10) models are left-right symmetric at the GUT scale
and we would expect similar mixing angles for quarks and leptons of both chiralities.
In left-right symmetric models smallness of left mixings implies that also right-handed
mixings are small, so that all mixings tend to be small, unless 
non-renormalizable mass operators with a suitable flavour pattern are 
introduced. In such a context, accommodating the observed mixing properties of
quarks and leptons appears more problematic than in SU(5) theories, 
at first sight. 

One possibility is to exploit the see-saw mechanism to enhance 
the light neutrino mixing angles. As illustrated in eq. (\ref{md1}) in a 2 by 2 context,
in order to have large or even maximal mixing in $m_\nu$, we do not necessarily need
large mixing angles in $m_D$ and $M$. We can obtain a large mixing
starting from nearly diagonal mass matrices $m_D$ and $M$, provided a RH neutrino,
equally coupled to $\nu_\mu$ and $\nu_\tau$, is sufficiently light \cite{lfsmall,chenm,dorsmi}.

An other possibility is to argue that perhaps what appears to be large is not that large after all. 
The typical small parameter that appears in the mass matrices is $\lambda\sim
\sqrt{m_d/m_s}
\sim
\sqrt{m_{\mu}/m_{\tau}}\sim 0.20-0.25$. This small parameter is not so small that it cannot become large due to some
peculiar accidental enhancement: either a coefficient of order 3, or an exponent of the mass ratio which is less
than $1/2$ (due for example to a suitable charge assignment), or the addition in phase of an angle from the
diagonalization of charged leptons and an angle from neutrino mixing.  
Typically, by exploiting the freedom in the parameter space, in this set of models
a large $\theta_{23}$ may be accommodated.
The large mixing angle for solar neutrinos requires however the introduction
of ad hoc terms, like for instance higher-dimensional operators 
contributing to the light neutrino masses independently 
from the see-saw mechanism \cite{bpw,str,pati}.  

Alternatively, to avoid the introduction of ad hoc non-renormalizable operators, it is possible to enlarge 
the Higgs content and consider, for example, an SO(10) model where all fermion mass matrices originate
from renormalizable interactions of matter fields in three 16 representations with two Higgs multiplets, 
a $10_H$ and a $126_H$
\cite{ten126}:
\beq
{\cal L}_Y= 10_H~ 16~ y_{10} 16 + 126_H~ 16~ y_{126} 16~~~~~,
\eeq
where $y_{10}$ and $y_{126}$ are two symmetric matrices in flavour space.
Both $y_{10}$ and $y_{126}$ contribute to the Dirac mass matrices,
with the characteristic factor -3 for the $y_{126}$ between the $(u,d)$ and $(\nu,e)$ sectors:
\bea
m_d&=&\alpha~ y_{10}+\beta~ y_{126}\nn\\
m_e&=&\alpha~ y_{10}-3\beta~ y_{126}~~~~~,
\label{mdme}
\eea
and the correct mass relations for first and second generations can be accommodated.
In the most general case, when $126_H$ acquires VEVs in its SU(2)$_L$ singlet and triplet components,
both RH and LH Majorana masses can arise in the neutrino sector. They are both proportional to $y_{126}$
and give rise to type I and type II see-saw, respectively. 
By assuming dominance of the type II contribution, we find an interesting link
between large atmospheric mixing angle and $b-\tau$ unification \cite{baj}.
In this case the light neutrino mass matrix is proportional to $y_{126}$ and, from eq. (\ref{mdme}), it 
can be directly related to $m_e$ and $m_d$:
\beq
m_\nu\propto m_d-m_e~~~~~.
\eeq
If both $m_e$ and $m_d$ have the approximate pattern
\beq
\left(
\matrix{
\lambda^2& \lambda^2\cr
\lambda^2& 1}
\right)~~~~~
\eeq
in the 23 sector, then $b-\tau$ unification forces a cancellation in the 33 entry of $m_\nu$,
thus allowing for a large 23 neutrino mixing angle. 
Most of the Yukawa parameters can be determined by the quark masses and mixing angles and from the charged
lepton masses. This allows to predict the atmospheric and solar mixing parameters within a range which is still 
experimentally allowed and $\vert U_{e3}\vert\approx 0.16$, not far from the present upper bound
\cite{moha}.
A drawback of the model is the occurrence of two pairs of Higgs doublets with non-vanishing VEVs,
of which only one combination is allowed to remain light. This clearly makes the doublet-triplet splitting 
problem even more complicated than in minimal models.

Finally, we can abandon the idea that the model is left-right symmetric at the GUT scale.
In this case, the mechanism discussed in section 8, based on asymmetric mass
matrices, can be embedded in an
SO(10) grand-unified theory in a rather economic way
\cite{barr,lops1,bpw,soten,albarr}. The 33 entries of the fermion mass matrices can be obtained through the coupling
${\bf 16}_3 {\bf 16}_3 {\bf 10}_H$ among the fermions in the third generation, ${\bf 16}_3$, and a Higgs tenplet
${\bf 10}_H$. The two independent VEVs of the tenplet $v_u$ and $v_d$ give mass, respectively, to $t/\nu_\tau$ and
$b/\tau$. The key point to obtain an asymmetric texture is the introduction of an operator of the kind ${\bf 16}_2
{\bf 16}_H {\bf 16}_3 {\bf 16}_H'$ . This operator is thought to arise by integrating out an heavy {\bf 10} that
couples both to ${\bf 16}_2 {\bf 16}_H$ and to ${\bf 16}_3 {\bf 16}_H'$. If the ${\bf 16}_H$ develops a VEV breaking
SO(10) down to SU(5) at a large scale, then, in terms of
SU(5) representations, we get an effective coupling of the kind ${\bf \bar{5}}_2 {\bf 10}_3 {\bf\bar{5}}_H$, with
a coefficient that can be of order one. This coupling contributes to the 23 entry of the down quark mass matrix  and
to the 32 entry of the charged lepton mass matrix, realizing the desired asymmetry.   To distinguish the lepton and
quark sectors one can further introduce  an operator of the form ${\bf 16}_i {\bf 16}_j {\bf 10}_H {\bf 45}_H$,
$(i,j=2,3)$, with the VEV of the 
${\bf 45}_H$ pointing in the $B-L$ direction. Additional operators, still of the type 
${\bf 16}_i {\bf 16}_j {\bf 16}_H {\bf 16}_H'$ can contribute to the matrix elements of the first generation. The
mass matrices look like:
\beq 
m_u=
\left(
\matrix{ \eta& 0& 0\cr 0& 0& \epsilon/3\cr 0&-\epsilon/3&1}
\right)v_u~~,~~~~~~~ m_d=
\left(
\matrix{ 0&\delta&\delta'\cr
\delta&0&\sigma+\epsilon/3\cr
\delta'&-\epsilon/3&1}
\right)v_d~~,
\label{mquark1}
\eeq 
\beq m_D=
\left(
\matrix{ \eta& 0& 0\cr 0& 0& -\epsilon\cr 0& \epsilon&1}
\right)v_u~~,~~~~~~~ m_l=
\left(
\matrix{ 0&\delta&\delta'\cr
\delta&0&-\epsilon\cr
\delta'&\sigma+\epsilon&1}
\right)v_d~~,
\label{mquark2}
\eeq  
\beq M=
\left(
\matrix{ b^2 \eta^2& -b \epsilon\eta& a \eta\cr -b \epsilon \eta& \epsilon^2& -\epsilon\cr a \eta & -\epsilon&1}
\right)\Lambda~~,
\eeq  
where $\eta\ll\delta,\delta'\ll\epsilon\ll 1$ and $a$, $b$ and $\sigma$ are of order O(1).
In the charged fermion sector, the parameter $\eta$, $\delta$, $\delta'$, $\epsilon$ and $\sigma$ 
are determined from the lepton masses and from a subset of quark masses and mixing angles,
leading to six successful predictions. In the neutrino sector, the lopsidedness of $m_l$
is responsible for the large atmospheric mixing angle, whereas the parameters $a$ and $b$ can be
adjusted to obtain the solar mixing angle and the ratio between solar and atmospheric squared mass 
differences. The model predicts small values for $\vert U_{e3}\vert$, in a range accessible only 
to future neutrino factories \cite{albarfit}.

Models based on SO(10) times a flavour symmetry are more difficult to construct because a whole generation is
contained in the 16, so that, for example for U(1)$_{\rm F}$, one would have the same value of the charge for 
all quarks and leptons of each generation, which is too rigid. This problem can be circumvented if not all the 
observed fermions in a given generation belong to a single 16 representation \cite{asaka}.  

\section{Conclusion}

By now there are rather convincing experimental indications for neutrino oscillations.  The direct implication of these
findings is that neutrino masses are not all vanishing. As a consequence, the phenomenology of neutrino masses and mixings
is brought to the forefront.  This is a very interesting subject in many respects. It is a window on the physics of GUTs in
that the extreme smallness of neutrino masses can only be explained in a natural way if lepton number conservation is
violated.  If so, neutrino masses are inversely proportional to the large scale where lepton number is violated. Also, the
pattern of neutrino masses and mixings interpreted in a GUT framework can provide new clues on the long standing problem of
understanding the origin of the hierarchical structure of quark and lepton mass matrices. 

Neutrino oscillations only
determine differences of $m_i^2$ values and the actual scale of neutrino masses remain to be experimentally fixed. The
detection of
$0\nu\beta\beta$ decay would be extremely important for the determination of the overall scale of neutrino masses, the
confirmation of their Majorana nature and the experimental clarification of the ordering of levels in the associated
spectrum. The recent results from cosmology indicate that neutrino masses are not a major fraction of the cosmological
mass density $\Omega_{\nu}\lappeq 1.5\%$. The decay of heavy right-handed neutrinos with lepton number non-conservation
can provide a viable and attractive model of baryogenesis through leptogenesis. The measured oscillation frequencies and
mixings are remarkably consistent with this attractive possibility.

While the existence of oscillations  appears to be on a ground of increasing solidity, many important experimental
challenges remain. For atmospheric neutrino oscillations the completion of the K2K experiment, which was delayed by the accident
that has seriously damaged the Superkamiokande detector, is important for a terrestrial confirmation of the effect and for
an independent measurement of the associated parameters. In the near future the experimental study of atmospheric
neutrinos will be further pursued with long baseline measurements by MINOS, OPERA, ICARUS.  For solar neutrinos the
continuation of SNO, KamLAND and the data from Borexino will lead to a more precise determination of the parameters of
the LA solution.  Finally a clarification by MINIBOONE of the issue of the LSND alleged signal is necessary, in order to
know if 3 light neutrinos are sufficient or additional sterile neutrinos must be introduced, in spite of the apparent lack
of independent evidence in the data for such sterile neutrinos and of the fact that attempts of constructing plausible and
natural theoretical models have not led so far to compelling results. Further in the future there are projects for neutrino
factories and/or superbeams aimed at precision measurements of the oscillation parameters and possibly the detection of CP
violation effects in the neutrino sector.
%\marginpar{aggiungere informazioni future da galaxies, MBR anisotropies,...}

Pending the solution of the existing experimental ambiguities a variety of theoretical models of neutrino masses and
mixings are still conceivable. Among 3-neutrino models we have described a number of possibilities based on degenerate,
inverted hierarchy and normal hierarchy type of spectra. The normal hierarchy option appears to us as the most
straightforward and flexible framework. In particular the large atmospheric mixing can arise from lopsided matrices. Then the
observed frequencies and the large solar angle can also be obtained without fine tuning in models where the 23 subdeterminant
is automatically suppressed. 

The fact that some neutrino mixing angles are large and even
nearly maximal, while surprising at the start, was eventually found to be well compatible with a unified picture of quark
and lepton masses within GUTs. The symmetry group at
$M_{GUT}$ could be either (SUSY) SU(5) or SO(10)  or a larger group. For example, we
have seen that models based on anarchy, semianarchy, inverted hierarchy or normal hierarchy can all be naturally
implemented  by simple assignments of U(1)$_{\rm F}$ horizontal charges in a semiquantitative unified
description of all quark and lepton masses in SUSY SU(5)$\times$ U(1)$_{\rm F}$. Actually, in this context, if one adopts
a statistical criterium, hierarchical models appear to be preferred over anarchy and among them normal hierarchy appears the most likely. Note that in almost all of the existing models of neutrino mixings the 
atmospheric angle is large but not  maximal. If it experimentally 
turned out that indeed $\theta_{23}$ is maximal with good accuracy then 
very special classes of models would be selected \cite{bitri,wett,maxatm}.

All we know about neutrino masses is well in harmony with the idea and the mass scale of GUT's. As a consequence neutrino
masses have added phenomenological support to this beautiful idea and to the models of physics beyond the Standard Model
that are compatible with it. In particular, if we consider the main classes of new physics that are currently
contemplated, like supersymmetry, technicolour, large extra dimensions at the TeV scale, little Higgs models etc, it is
clear that the first one is the most directly related to GUT's. SUSY offers a well defined model computable up to the GUT
scale and is actually supported by the quantitative success of coupling unification in SUSY GUT's. For the other examples
quoted all contact with GUT's is lost or at least is much more remote. In this sense neutrino masses fit particularly well
in the SUSY picture that sofar remains the standard way beyond the Standard Model. 
\vspace{1.0cm}
\section{Acknowledgements} F.F. is partially supported by the European Programs HPRN-CT-2000-00148
and HPRN-CT-2000-00149.
\vfill
\newpage
%
%\vspace*{3.0cm}

%
\end{document}